\documentclass[twocolumn,preprintnumbers,amsmath,amssymb,aps,prb,longbibliography]{revtex4-1}
\usepackage{graphicx}
\begin{document}

\title{Shapiro Steps and Nonlinear Skyrmion Hall Angles For dc and ac Driven Skyrmions on a Two Dimensional
Periodic Substrate }
 
\author{N. P. Vizarim$^{1,2}$, C. Reichhardt$^{1}$, P. A. Venegas$^3$, and C. J. O. Reichhardt$^{1}$}
\affiliation{$^{1}$Theoretical Division and Center for Nonlinear Studies,
  Los Alamos National Laboratory, Los Alamos, New Mexico 87545, USA}
\affiliation{$^2$ POSMAT - Programa de P{\' o}s-Gradua{\c c}{\~ a}o em Ci{\^ e}ncia e Tecnologia de Materiais, Faculdade de Ci\^{e}ncias, Universidade Estadual Paulista - UNESP, Bauru, SP, CP 473, 17033-360, Brazil}
\affiliation{$^{3}$ Departamento de F\'{i}sica, Faculdade de Ci\^{e}ncias, Universidade Estadual Paulista - UNESP, Bauru, SP, CP 473, 17033-360, Brazil
}

\date{\today}

\begin{abstract}
For an overdamped particle moving over a two-dimensional
periodic substrate under combined dc and ac drives, 
a series of steps can appear in the velocity force curves that are known
as Shapiro steps. Here we show that for
skyrmions driven over a two-dimensional periodic obstacle array with
a dc drive and an ac drive that is either parallel or perpendicular to the dc
drive,
the system exhibits
numerous transverse and longitudinal synchronization dynamics due
to the Magnus force.
These phenomena
originate in interactions between two different types of phase locking effects:
Shapiro steps and directional locking.
In some cases, the skyrmion Hall angle is constant but
longitudinal Shapiro steps appear, while in other regimes
the skyrmion Hall angle can either increase or decrease with
increasing dc drive during the phase locking
as the skyrmion locks to different symmetry directions of the obstacle lattice.
For a transverse ac drive we find that
strong Hall angle overshoots can occur in certain locked phases
where the skyrmion is moving at an angle that is considerably
larger than the intrinsic Hall angle.
For the strongest Magnus force, the phase locking effects are reduced and
there are larger regions of disordered dynamics.
We show that the skyrmion Hall angle can be controlled by fixing the dc drive
and changing the amplitude of the ac drive.  
\end{abstract}

\maketitle

\vskip 2pc

\section{Introduction}
Systems with multiple interacting frequencies are known to exhibit various
nonlinear dynamical effects
such as synchronization or phase locking \cite{Pikovsky01,Ott93}. 
Such phenomena arise across a wide range of fields
ranging from coupled pendula \cite{Bennett02} to biological systems \cite{Glass01}.
One of the simplest examples of a system that 
can exhibit phase locking is an overdamped particle on
a periodic substrate under a combined dc and ac drive, 
where there can be resonances
between the ac driving frequency and the
frequency of the oscillations generated by the motion of the particle
over the periodic substrate.
These resonance effects create
a series of steps in the velocity force curves since
the particle remains locked to
a specific velocity over an interval of the external drive
in order to remain in the resonant state.
One of the first
systems where such resonant steps were observed was
Josephson junctions, where so-called Shapiro steps appear in the current-voltage 
response \cite{Shapiro63,Barone82}.
Many systems that exhibit phase locking can be described as effectively one
dimensional, and locking dynamics have been studied for 
Josephson junction arrays \cite{Benz90},
incommensurate sliding charge density waves \cite{Coppersmith86,Gruner88},
vortices in type-II superconductors with
one-dimensional (1D) \cite{Martinoli75,Martinoli78,Dobrovolskiy15}
and two-dimensional (2D) periodic substrates \cite{vanLook99,Reichhardt00b},
driven Frenkel-Kontorova models \cite{Sokolovic17}, frictional systems
\cite{Tekic11},
and colloids moving over 1D periodic substrates
\cite{Juniper15,Brazda17,Abbot19}.
Even in the 1D case, a variety of additional phenomena 
such as fractional locking
can arise when additional nonlinear effects come into play.

Particles moving over a periodic 2D substrate exhibit many of the same phase locking
effects as the 1D systems,
but the additional degrees of freedom available in 2D make it
possible to align the ac drive perpendicular to the dc drive.
In this case, new phase locking effects 
that are distinct from Shapiro steps can appear that are known as
transverse phase locking, in which the step widths generally grow with
increasing ac amplitude \cite{Reichhardt01,Marconi03}
rather than oscillating with increasing ac amplitude as in Shapiro steps.
For 2D substrates it is also possible
to have bi-harmonic ac drives applied both
parallel and perpendicular to the dc drive, which generate
a circular motion of the driven particle.
Here, an increasing dc drive produces
chiral scattering effects
that result in
phase locked regions
in which the particle motion is both
transverse and longitudinal to the dc drive direction
\cite{Reichhardt02,Reichhardt03}.  

In most of the above systems the dynamics is 
overdamped; however, in some situations non-dissipative effects
such as inertia can arise \cite{Tekic19}.
Another type 
of non-dissipative effect is a
gyro-coupling or Magnus force, which creates
velocity components that are
perpendicular to the forces experienced by the particle.
In a 1D system, a Magnus force has no effect;
however, in 2D systems it can strongly modify the dynamics.
Magnus forces can be significant or even dominating for
skyrmions in chiral magnets
\cite{Muhlbauer09,Yu10,Nagaosa13,Jiang17a},
where the ratio of the Magnus force to the damping term can vary from $0.1$ to $10$. 
Skyrmions can interact with pinning sites,
be set into motion readily with an applied 
current, and exhibit depinning thresholds
\cite{Nagaosa13,Schulz12,Iwasaki13,Lin13,Liang15,Woo16,Montoya18}. 
One of the most prominent effects of the Magnus force is that
the skyrmions move at an angle with respect to the applied
driving force which is known as the skyrmion Hall angle
$\theta_{sk}$ \cite{Nagaosa13}, as has been observed in
simulations \cite{Iwasaki13a,Reichhardt15a,Reichhardt15,Reichhardt19b}
and experiments \cite{Jiang17,Litzius17}. 
The Magnus force strongly modifies the interaction of the skyrmion
with a substrate 
by creating spiraling motions of skyrmions that are in a trapping potential
\cite{Litzius17,Liu13,Muller15,Buttner15,Martinez16,Navau16,Gong20}. 
The pinning or defects produce a strong drive dependence of the 
skyrmion Hall angle, which starts off near zero just at depinning
and increases with increasing skyrmion
velocity before saturating to the intrinsic value at high drives.
This effect was first observed in 
simulations of skyrmions interacting with periodic or random disorder
\cite{Reichhardt15a,Reichhardt15,Reichhardt19b,Muller15,Legrand17,Kim17}
and was then found in experiments
\cite{Jiang17,Litzius17,Woo18,Juge19,Zeissler20,Litzius20}.
The drive dependence arises
due to 
a side jump effect when the skyrmion scatters off a pinning
site \cite{Reichhardt15a,Muller15}.
For random disorder, $\theta_{sk}$ increases smoothly
with increasing drive; however, for 
a periodic substrate a guiding effect occurs which causes
the skyrmion motion to  become directionally locked
to specific symmetry directions of the substrate over a range of drives,
producing
a quantized skyrmion Hall angle \cite{Reichhardt15a,Feilhauer19}. 
There are a number of proposals on how to create localized skyrmion pinning sites,
which can be attractive or repulsive \cite{Stosic17,Fernandes18,Toscano19,Xiong19},
and there are now experimental realizations
of skyrmion phases in periodic substrates \cite{Saha19a}
and superconducting vortices interacting with skyrmions \cite{Palermo20},
making it feasible to create
tailored 2D pinning 
arrays of attractive or repulsive obstacles with which skyrmions can interact.

Skyrmions moving over a 2D substrate under dc and ac drives
are expected to exhibit a variety of new synchronization effects not observed
in overdamped systems.
In numerical work examining dc and ac driven skyrmions
in 2D systems moving over a periodic 1D substrate, Shapiro steps
appeared for the velocity in both the longitudinal and transverse
directions \cite{Reichhardt15b}.
In an overdamped 2D system with a 1D substrate,
phase locking occurs only when the dc drive, ac drive, and substrate periodicity direction
are all aligned.
The inclusion of the Magnus force
allows any combination of the ac and dc drive directions
to produce some form of phase locking, and also generates
new effects such as
Shapiro spikes in the velocity force curves,
which are distinct from Shapiro steps \cite{Reichhardt17}.
It is even possible for
absolute transverse mobility to appear in which the skyrmion moves
at $90^\circ$ to the driving direction, as well as
negative mobility in which the net skyrmion motion
is in the direction opposite to the applied 
drive \cite{Reichhardt17}, or ratchet effects \cite{Chen19,Chen20}.
The Magnus force opens entirely new aspects of nonlinear dynamics,
and skyrmions moving over periodic substrates
can serve to provide experimental realizations of such dynamics.
These results suggest that skyrmion motion can
be controlled by combining a periodic substrate array
with different
driving protocols, which can be important for applications \cite{Fert13}.

In this work we extend our previous results on dc and ac driven
skyrmions on a 1D periodic array \cite{Reichhardt15b,Reichhardt17,Reichhardt15aa}
to the case of skyrmions interacting with a 2D periodic
array of obstacles, where the ac drive can be applied either parallel or
perpendicular to the dc drive direction.
We find two dominant effects.
The first is directional locking,
which is similar to that found previously for purely dc driven skyrmions
on a 2D substrate \cite{Reichhardt15,Feilhauer19,Vizarim20}.
The second is
Shapiro steps similar to those found
for skyrmions and vortices under ac and dc drives on a 1D periodic substrate
\cite{Reichhardt15b}.
These two effects can interfere with each other.
In some cases, we find a constant
skyrmion Hall angle accompanied by
steps in both the parallel and perpendicular velocity components,
while in other cases, a series of steps in the skyrmion Hall angle
coexists with regimes in which the skyrmion Hall angle is non-monotonic and
either increases, decreases, or reverses sign as a function of the applied drive.
We also show that for fixed dc driving, the skyrmion Hall angle can
be controlled by changing the amplitude of the ac drive.

\section{Simulation}

\begin{figure}
\includegraphics[width=3.5in]{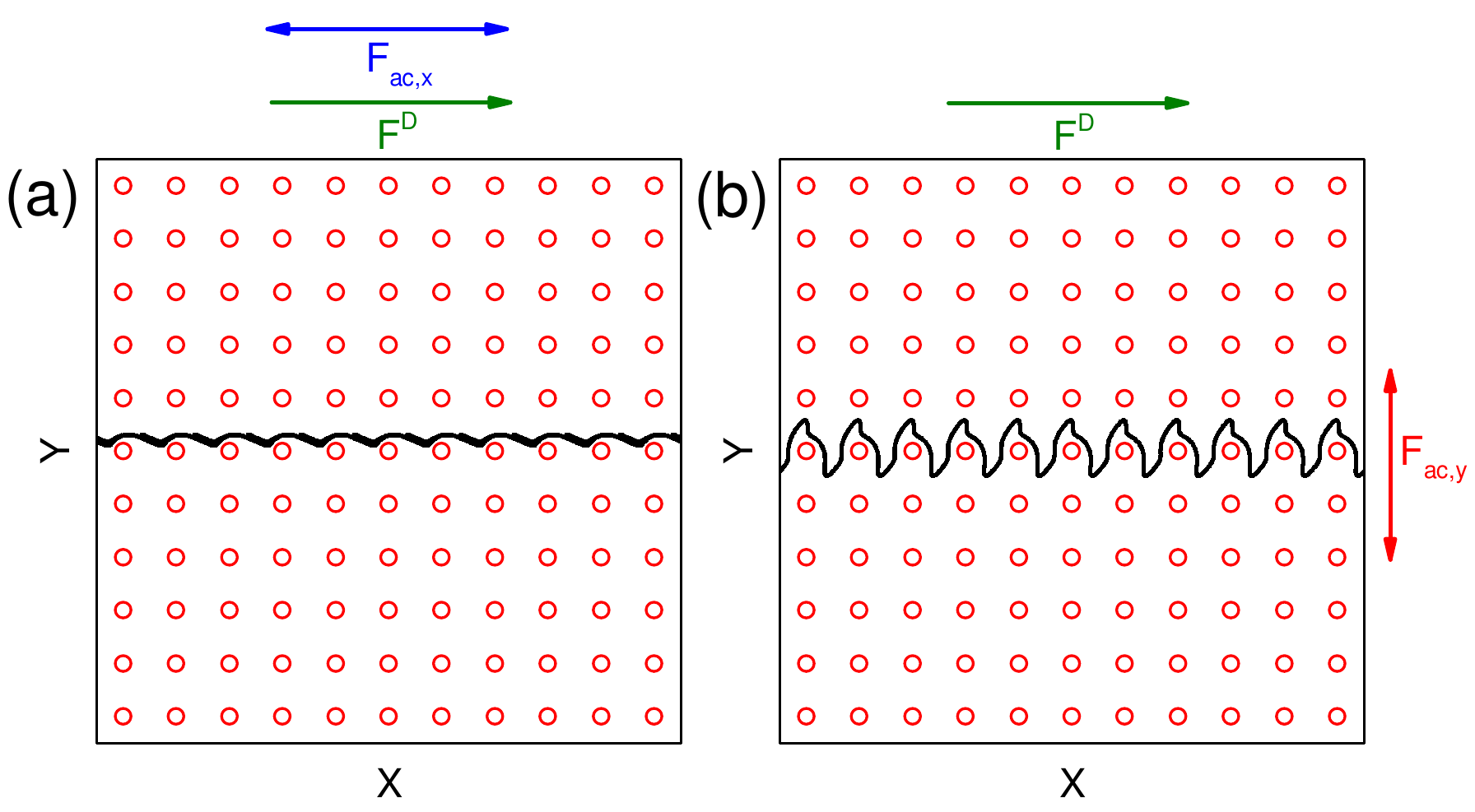}
\caption{ A schematic of the system, which consists of a square array of obstacles
  (red circles) modeled as repulsive Gaussian scattering sites.
  The black line is the trajectory of a skyrmion which is subjected to both damping and
  a Magnus term as well as a dc drive $F^D$ applied along the $x$ direction.
  An additional ac drive is applied either (a) along the $x$ direction, $F_{ac,x}$, or
  (b) along the $y$ direction, $F_{ac,y}$.}
\label{fig:1}
\end{figure}

We consider a two-dimensional $L \times L$ system with periodic boundary conditions
and model a single skyrmion 
moving over a square obstacle array with lattice constant $a$.
In Fig.~\ref{fig:1} we show a schematic of the system
highlighting the obstacles and the skyrmion trajectory
for a dc drives $F^{D}$ applied
along the $x$-direction.
We apply an additional ac drive 
along the $x$-direction, as shown in Fig.~\ref{fig:1}(a),
or along the $y$-direction, as illustrated in Fig.~\ref{fig:1}(b).
In the presence of only a dc drive,
a series of directional locking steps
appear due to the velocity dependence of the skyrmion Hall effect \cite{Vizarim20}.  
 
We use a particle based model for skyrmions interacting with disorder \cite{Lin13,Reichhardt15a,Reichhardt15,Vizarim20,Brown19}. 
The equation of motion for skyrmion $i$ is:
\begin{equation}
\alpha_d {\bf v}_{i} + \alpha_m {\hat z} \times {\bf v}_{i} =  {\bf F}^{obs} + {\bf F}^{D} + {\bf F}^{AC} 
\end{equation}
The first term $\alpha_{d}$  on the left is the damping term,
and the second term $\alpha_{m}$ is the Magnus force, 
which produces velocities that are perpendicular to the net force experienced
by the skyrmion.
Unless otherwise noted, we normalize the damping and Magnus
coefficients so that ${\alpha }^2_d+{\alpha }^2_m=1$.

The first term on the right, ${\bf F}^{obs}$,
represents the interaction between the skyrmion and the obstacles. 
The potential energy
of this interaction has a Gaussian form
$U_o=C_oe^{-{\left({r_{io}}/{a_o}\right)}^2}$, where $C_o$ 
is the strength of the obstacle potential, 
$r_{io}$ is the distance between skyrmion $i$ and obstacle $o$,
and $a_o$ is the obstacle radius. 
The force between an obstacle and the skyrmion is given by
${\boldsymbol{\mathrm{F}}}^o_i=-\mathrm{\nabla }U_o=-F_or_{io}e^{-{\left({r_{io}}/{a_o}\right)}^2}{\widehat{\boldsymbol{\mathrm{r}}}}_{io}\ $, 
where $F_o=2U_o/a^2_o$. 
For computational efficiency, we place a cut-off
on the obstacle interaction at $r_{io}=2.0$ since the interaction
becomes negligible beyond this
distance.
We set the obstacle density to
$\rho_o=0.093/{\xi }^2$ and the obstacle radius to $a_0=0.65$. 
The dc drive is represented by the
term ${\boldsymbol{\mathrm{F}}}^D\boldsymbol{=}F^D\widehat{\boldsymbol{\mathrm{x}}}$. 
We increase the dc drive in small increments of $\delta F^D=0.001$,
and we wait ${10}^5$ simulation time steps between increments to ensure
that the system has reached a steady state.
The ac drive has the form
${\bf F}^{AC} = A\sin(\omega t){\bf \hat{\alpha}}$ for
driving in the $x$ ($\hat{\alpha}={\hat {\bf x}}$) or
$y$ ($\hat{\alpha}={\hat {\bf y}}$) direction.   
Here $A$ is the amplitude
of the ac drive and the drive frequency is
$\omega=2 \times 10^{-4}$ inverse simulation steps.
We measure the 
skyrmion velocity parallel,
$\left\langle V_{\parallel }\right\rangle =\langle {\bf v}_i \cdot {\bf \hat x}\rangle$,
and perpendicular,
$\left\langle V_{\bot }\right\rangle=\langle {\bf v}_i \cdot {\bf \hat y}\rangle $,
to the dc drive.
In the absence of any obstacles, the 
skyrmion moves at the intrinsic skyrmion Hall angle,
$\theta_{sk}^{int}=
{\mathrm{arctan} \left({\alpha }_m/{\alpha }_d\right)\ }$.
We can also quantify the dynamics by measuring
$R=\ \left\langle V_{\bot }\right\rangle /\left\langle V_{\parallel }\right\rangle$, where the
skyrmion Hall angle is given by $\theta_{sk} = \arctan(R)$.

\section{DC and AC Drive in the Same Direction}

\begin{figure}
\includegraphics[width=3.5in]{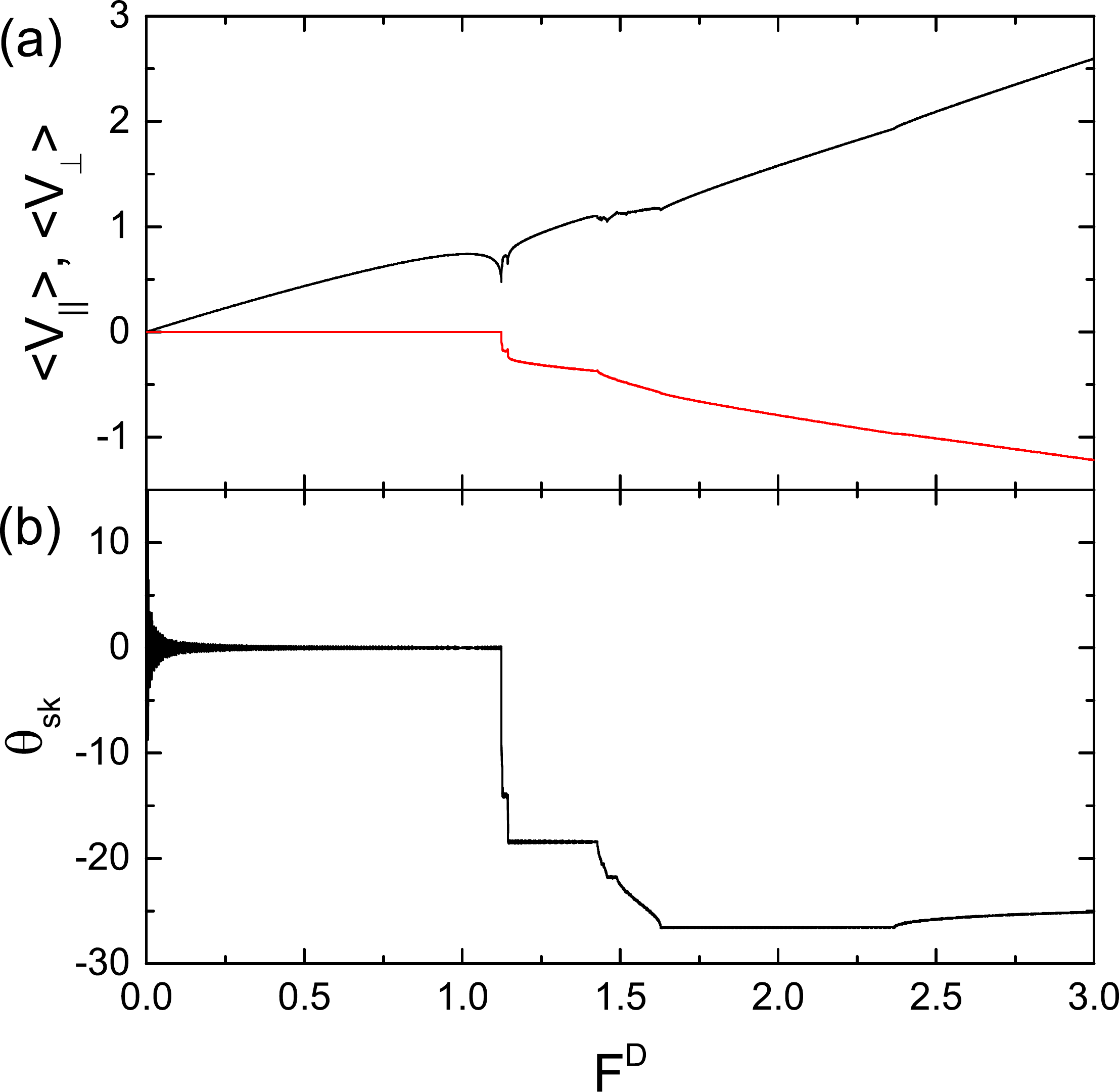}
\caption{  
$\langle V_{\perp}\rangle$ (red) and $\langle V_{||}\rangle$ (black)
vs $F^{D}$ for strictly dc driving with an ac drive amplitude of
$A = 0.0$ in a sample with
$\alpha_{m}/\alpha_{d} = 0.45$.
(b) The corresponding $\theta_{sk}$ vs $F^D$.}
\label{fig:2}
\end{figure}

We first consider the case where
the ac drive is applied along the same direction as
the dc drive, ${\bf F}^{AC}=A\sin(\omega t){\bf \hat{x}}$.
For an overdamped particle
moving over a periodic array,
this drive configuration produces Shapiro steps in the
velocity-force curves, and the motion is strictly in the
drive direction, giving a Hall angle of zero. 
In Fig.~\ref{fig:2}(a) we plot
$\langle V_{\perp}\rangle$ and $\langle V_{||}\rangle$
versus $F^{D}$ at zero ac driving,
$A = 0.0$, in a sample with
$\alpha_{m}/\alpha_{d} = 0.45$,
while in Fig.~\ref{fig:2}(b) we show the corresponding
$\theta_{sk}$ versus $F^D$ curve.
A series of jumps in the velocity-force curves
are associated with different locking directions for the skyrmion
motion, as indicated by
the jumps in $\theta_{sk}$.
This is a result of the
pinning-induced
velocity dependence of the skyrmion Hall angle,
as previously studied for skyrmions moving over a periodic
pinning or obstacle array
\cite{Reichhardt15a,Vizarim20}.

\begin{figure}
\includegraphics[width=3.5in]{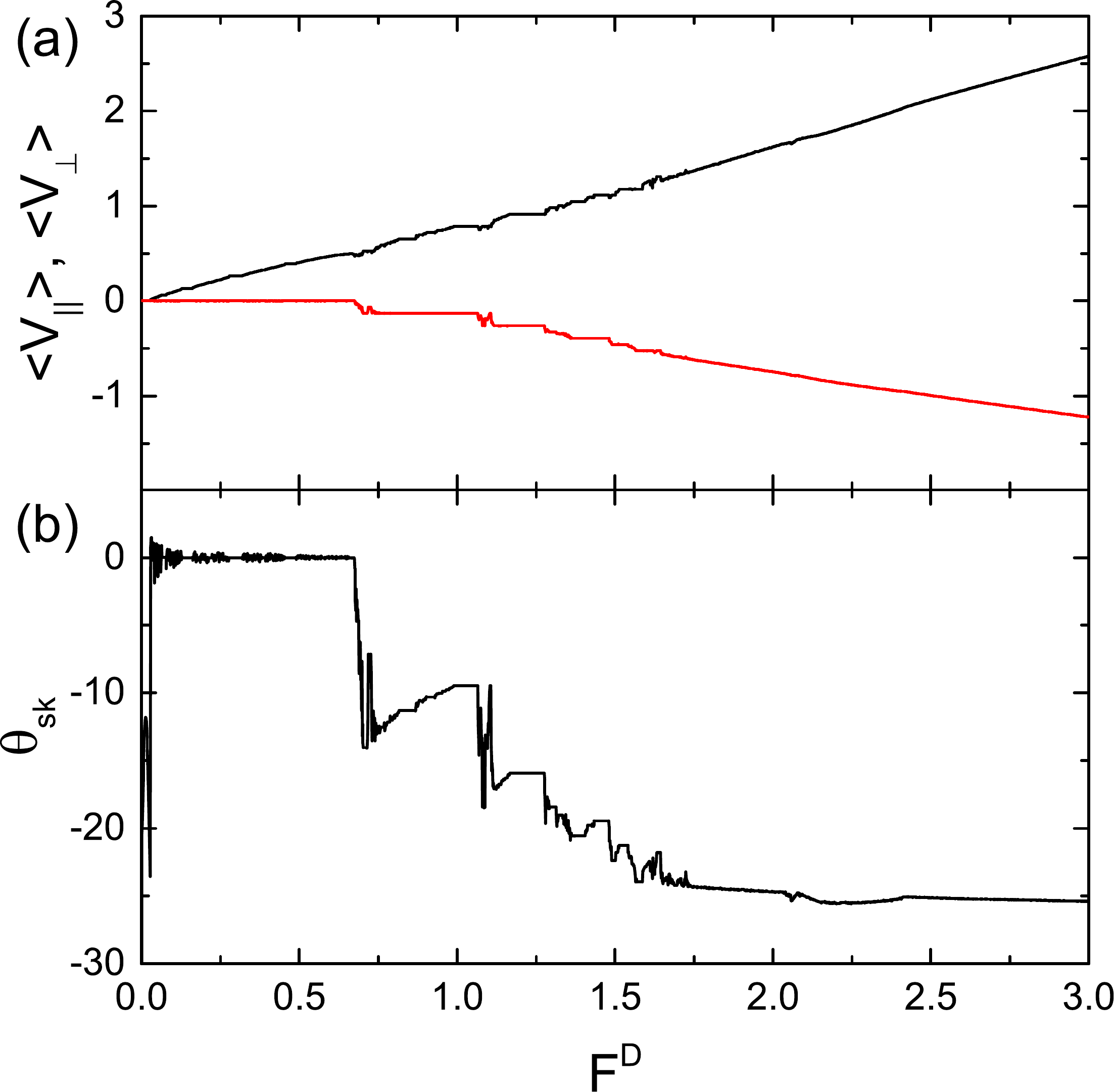}
\caption{
  (a) $\langle V_{||}\rangle$ (black)  and $\langle V_{\perp}\rangle$ (red) vs $F^{D}$
  for the system in Fig.~\ref{fig:2} with $\alpha_m/\alpha_d=0.45$
  under a finite ac drive with $A=0.5$ applied along the $x$ direction.
(b) The corresponding $\theta_{sk}$ vs $F_{D}$.
  There are windows of drive over which the magnitude of
  $\theta_{sk}$ decreases with increasing $F^{D}$.
}
\label{fig:3}
\end{figure}

When a finite ac drive of $A=0.5$ is applied along the $x$ direction in the same
system, the behavior changes as illustrated in Fig.~\ref{fig:3}.
In Fig.~\ref{fig:3}(a) we plot $\langle V_{||}\rangle$ and
$\langle V_{\perp}\rangle$ versus $F^{D}$,  while
in Fig.~\ref{fig:3}(b) we show the corresponding
$\theta_{sk}$ versus $F_{D}$ curve.
The skyrmion motion is initially locked along the
$x$-direction for $F^{D} < 0.65$,
and above this drive $\langle V_{\perp}\rangle$ begins
to increase in a series of steps. 
The skyrmion Hall
is non-monotonic between the steps.
Above the first step in
$\langle V_{\perp}\rangle$, the Hall angle is close to
$\theta_{sk} = -12.5^\circ$,
and it decreases in magnitude
with increasing drive to
$\theta_{sk} = -8^\circ$ before increasing
in magnitude again.
This pattern repeats several times until,
at high drives, $\theta_{sk}$ saturates
to $\theta_{sk} = 24^\circ$, a value close 
to the pin-free intrinsic skyrmion Hall angle.
At the higher drives, the steps in the velocity force curves
also become smoother. 
The decreases in magnitude of the skyrmion Hall angle
with increasing $F^{D}$ have not been observed
for skyrmions interacting
with random pinning.

\begin{figure}
\includegraphics[width=3.5in]{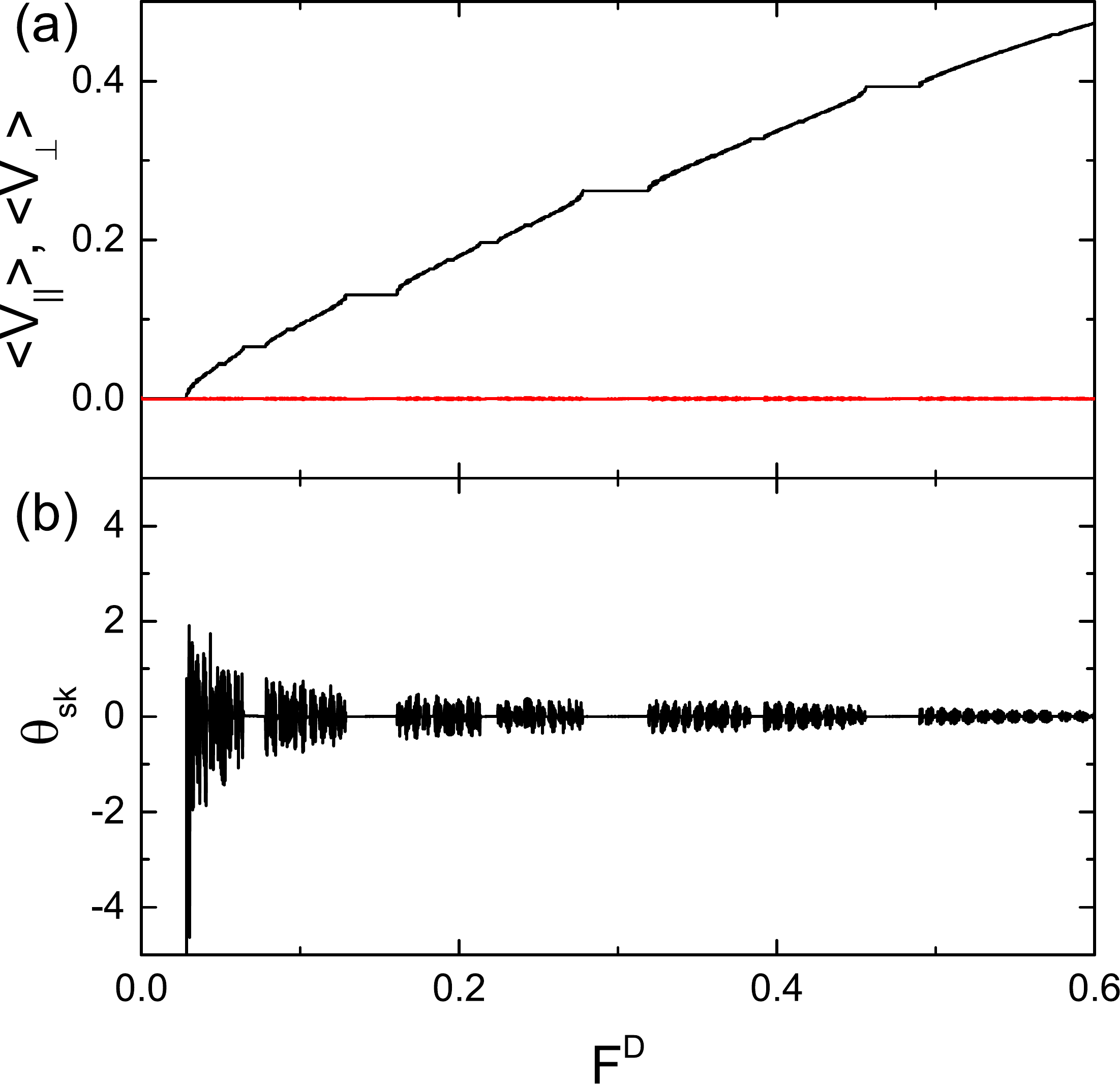}
\caption{ (a) $\langle V_{||}\rangle$ (black) and $\langle V_{\perp}\rangle$ (red)
  vs $F_{D}$ for the system in Fig.~\ref{fig:3} with $\alpha_m/\alpha_d=0.45$
  and $x$ direction ac driving of $A=0.5$ 
  zoomed in on the range $0.0 < F_{D} < 0.6$. 
(b) The corresponding $\theta_{sk}$ vs $F_{D}$ showing Shapiro steps. }
\label{fig:4}
\end{figure}

In Fig.~\ref{fig:4}(a,b),
we zoom in on the range $0<F_D<0.6$ 
for the two velocity components and $\theta_{sk}$ in the system
from Fig.~\ref{fig:3}.
Here
$\theta_{sk} = 0.0$ and
$\langle V_{\perp}\rangle = 0$, indicating that the motion is locked
along $x$ direction; however,
a set of phase locking steps still appear in
$\langle V_{||}\rangle$.
These are Shapiro steps, which also occur
in the overdamped limit.
The steps correspond to windows of drive over which
$\langle V_{||}\rangle$ is locked to a constant value.
In contrast,
the directional locking found in the absence of an ac drive
in
Fig.~\ref{fig:2} is not associated with constant velocity steps but instead
is accompanied by
dips and cusps in the velocities.

\begin{figure}
\includegraphics[width=3.5in]{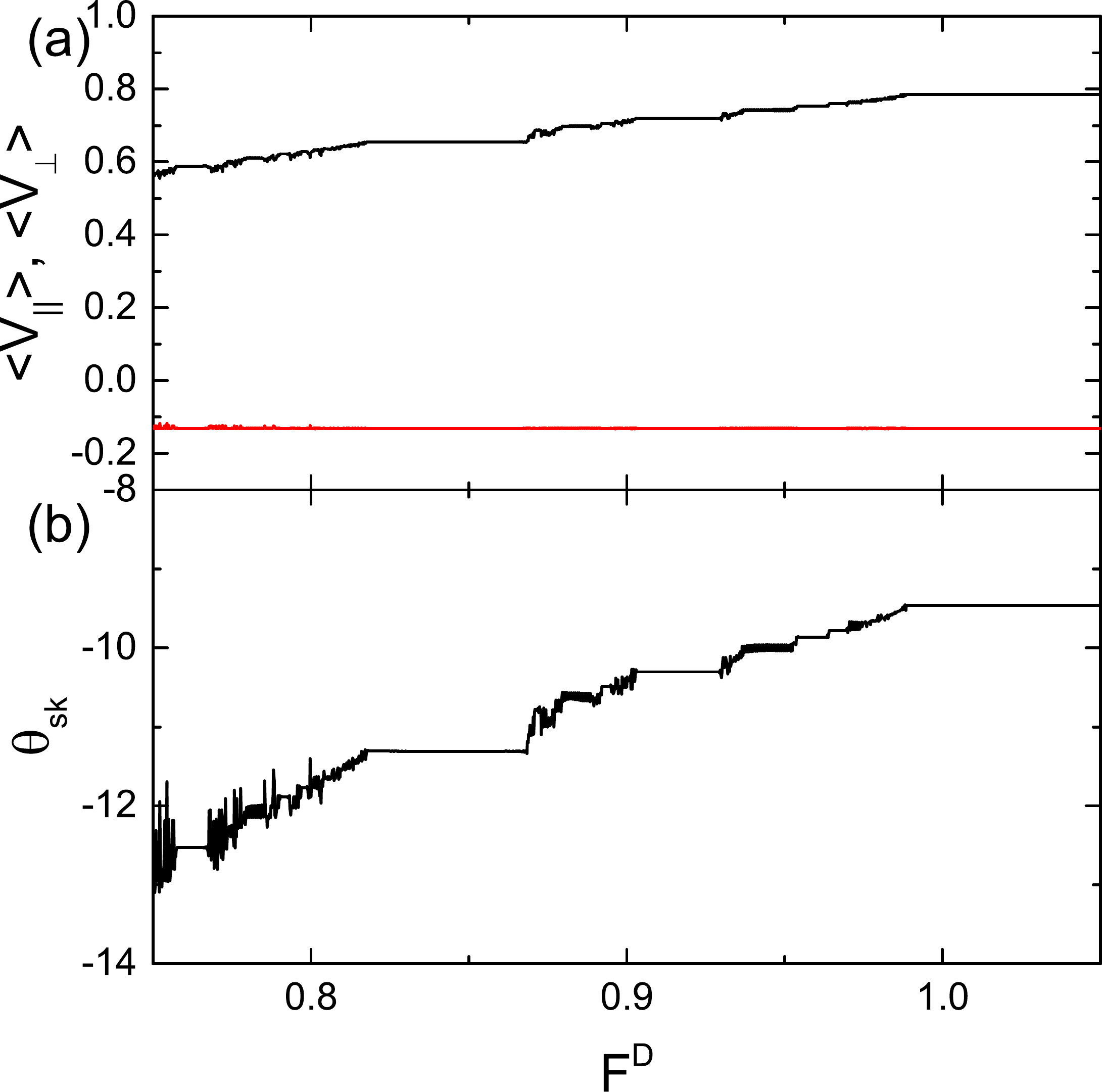}
\caption{ (a) $\langle V_{||}\rangle$ (black) and $\langle V_{\perp}\rangle$ (red)
  vs $F_{D}$ for the system in Fig.~\ref{fig:3} with $\alpha_m/\alpha_d=0.45$
  and $x$ direction ac driving of $A=0.5$ zoomed in on the
  range $0.7 < F_{D} < 1.2$.
  (b) The corresponding $\theta_{sk}$ vs $F_{D}$,
  showing a regime of decreasing skyrmion Hall angle magnitude.}
\label{fig:5}
\end{figure}

Figure~\ref{fig:5}(a,b) shows the curves from Fig.~\ref{fig:3}
over the interval
$0.7 < F^{D} < 1.2$, where we find two new features.
The first is that
$\langle V_{\perp}\rangle$ has a fixed finite value,
indicating that the particle is moving 
at an angle to the dc drive.
The second is that
the series of steps which appear in $\langle V_{||}\rangle$
are correlated with steps in $\theta_{sk}$, which is {\it decreasing} in magnitude
as $F^D$ increases.
This indicates that the velocity is
increasing in the $x$ direction but remains constant in the $y$ direction,
and the different phase locking steps are associated with decreases in the
magnitude of the skyrmion Hall angle.
Near $F^{D} = 1.2$ in Fig.~\ref{fig:3}, there is a substantial jump
in $\theta_{sk}$ to a larger magnitude which coincides with
a jump to a new step in $\langle V_{\perp}\rangle$.

The results in Figs.~\ref{fig:3}, \ref{fig:4}, and \ref{fig:5} show
that the phase locking behavior found in
Fig.~\ref{fig:3} is actually a {\it mixture} of two different types of locking.
The first
is the Shapiro step phase locking associated
with the matching of the ac drive frequency or its higher harmonics to
the increasing frequency of the skyrmion velocity oscillations caused by
the periodic collisions with the obstacles under an increasing dc drive.
This locking is associated with a $\theta_{sk}$ value that is either
constant or increasing in magnitude.
The second is the directional locking which occurs even in the absence
of an ac drive, as shown in Fig.~\ref{fig:2} and observed
in previous works \cite{Reichhardt15a,Feilhauer19,Vizarim20}.
These two locking phenomena can interact with each other to create
regions where the magnitude of the skyrmion Hall angle
is either constant or decreasing with drive instead of increasing with drive.
We note that directional locking effects for a particle 
moving over a periodic substrate can also occur
for overdamped systems such as vortices in type-II superconductors moving 
over 2D pinning arrays \cite{Reichhardt99,Reichhardt12}
and colloids moving over optical traps 
\cite{Reichhardt12,Korda02,MacDonald03,Lacasta05,Speer09,Balvin09}
or periodic substrates \cite{Cao19,Stoop20};
however, in those systems the direction of the
drive with respect to the substrate must be varied, whereas for the
skyrmions, 
the velocity dependence of the skyrmion Hall angle changes the
direction of motion even when the driving direction is fixed
\cite{Reichhardt15a,Vizarim20}.

\begin{figure}
\includegraphics[width=3.5in]{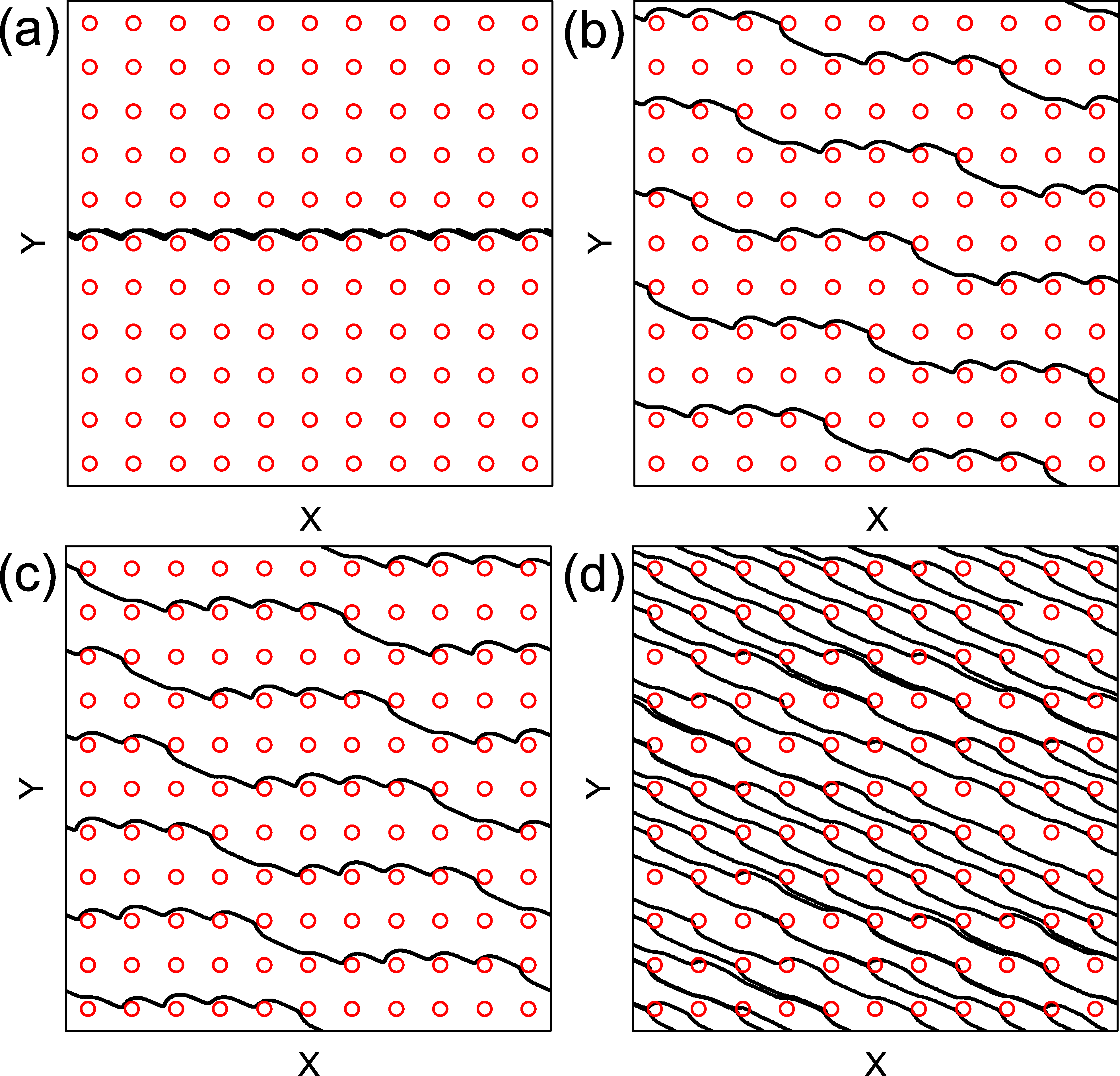}
\caption{ Skyrmion trajectory (black line) and obstacle locations (red circles)
  for the system in Fig.~\ref{fig:3} with $\alpha_m/\alpha_d=0.45$
  and
  $x$ direction ac driving with $A=0.5$.
  (a) $F^{D} = 0.3$ where the motion is locked in the $x$ direction.
  (b) $F^{D} = 0.85$ where there is finite motion along $y$ with $\theta_{sk} = -11.3^\circ$. 
  (c) $F^{D} = 1.0$, where $\theta_{sk} = -8.1^\circ$.
  (b) $F_{D} = 2.0$, where $\theta_{sk} = -22^\circ$.  
  }
\label{fig:6}
\end{figure}

In Fig.~\ref{fig:6}(a) we plot the skyrmion trajectories for the system in
Fig.~\ref{fig:3} at $F^{D} = 0.3$ where the skyrmion
motion is locked in the $x$-direction with $\theta_{sk} = 0^\circ$.
At $F^{D} = 0.85$ in Fig.~\ref{fig:6}(b),
the skyrmion has a finite displacement in the negative $y$-direction
and it traverses five obstacles in the $x$-direction for every one
obstacle in the $y$ direction, giving a ratio of 
$R = 1/5$
and $\theta_{sk} =\arctan(1/5) = -11.3^\circ$.
In Fig.~\ref{fig:6}(c) at $F^{D} = 1.0$,
the velocity in the $y$-direction is unchanged but the skyrmion
Hall angle has a smaller magnitude
of $-8.1^\circ$, and the skyrmion moves 7 lattice constants in $x$ and one
lattice constant in $y$ 
during a single ac drive period.
Figure~\ref{fig:6}(d) shows the trajectories in the same system
at $F^{D} = 2.0$, where $\theta_{sk}$ is 
close to $\theta_{sk}=-22^\circ$.
Here the system is not on a locking step and the trajectories are more disordered.

\begin{figure}
\includegraphics[width=3.5in]{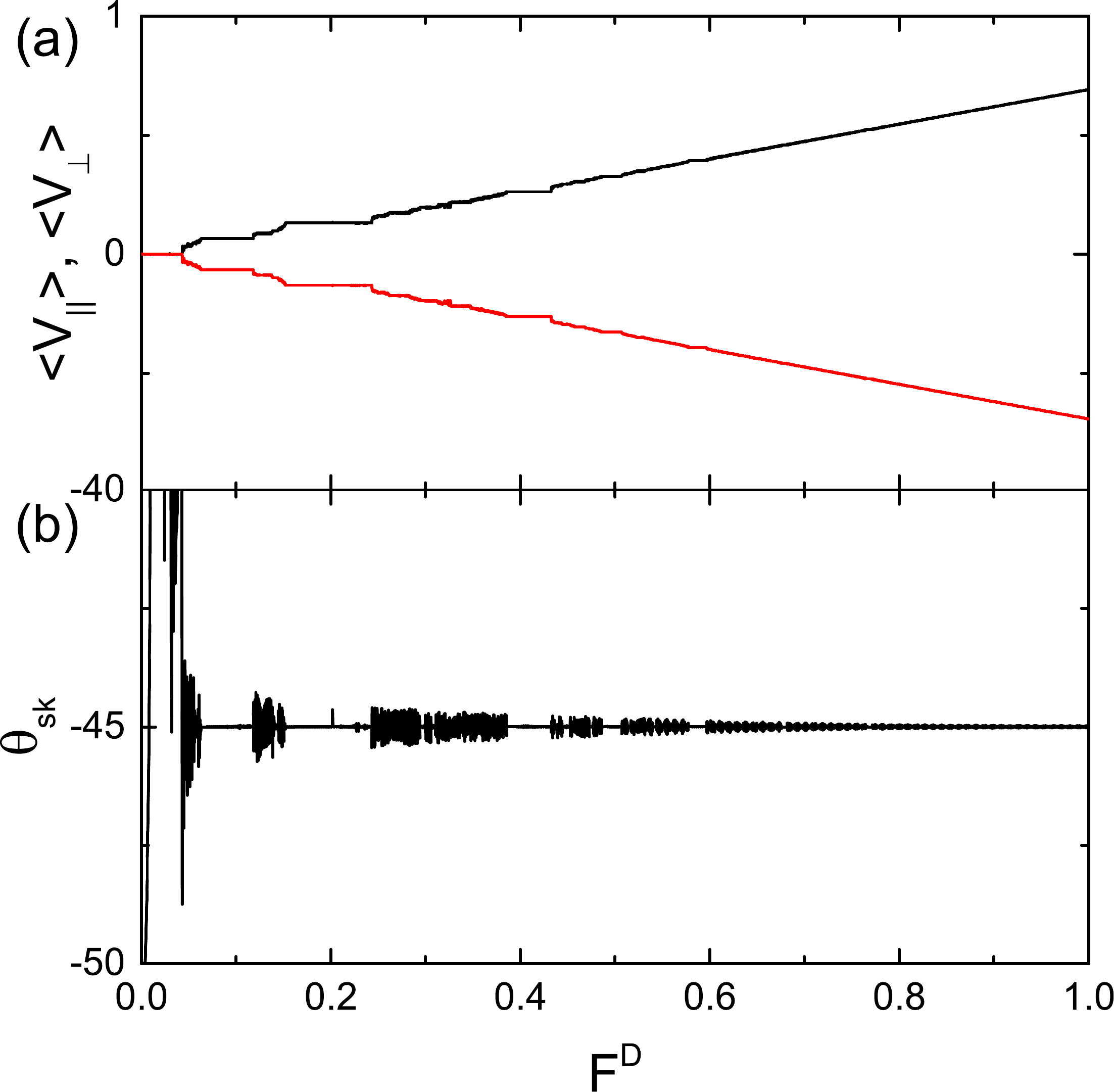}
\caption{ 
(a) $\langle V_{||}\rangle$ (black) and $\langle V_{\perp}\rangle$ (red)
  vs $F_{D}$ for a system with
  $x$ direction ac driving of magnitude
  $A=0.5$ at $\alpha_{m}/\alpha_{d} =1.0$, 
  where the intrinsic Hall angle is $\theta_{sk}^{int}=45^\circ$.
  (b) The corresponding $\theta_{sk}$ vs $F_{D}$.
  Here the motion is locked to $45^\circ$ but there
  is still a series of Shapiro steps that are
  not associated with a changing skyrmion Hall angle.    
}
\label{fig:7}
\end{figure}

\begin{figure}
\includegraphics[width=3.5in]{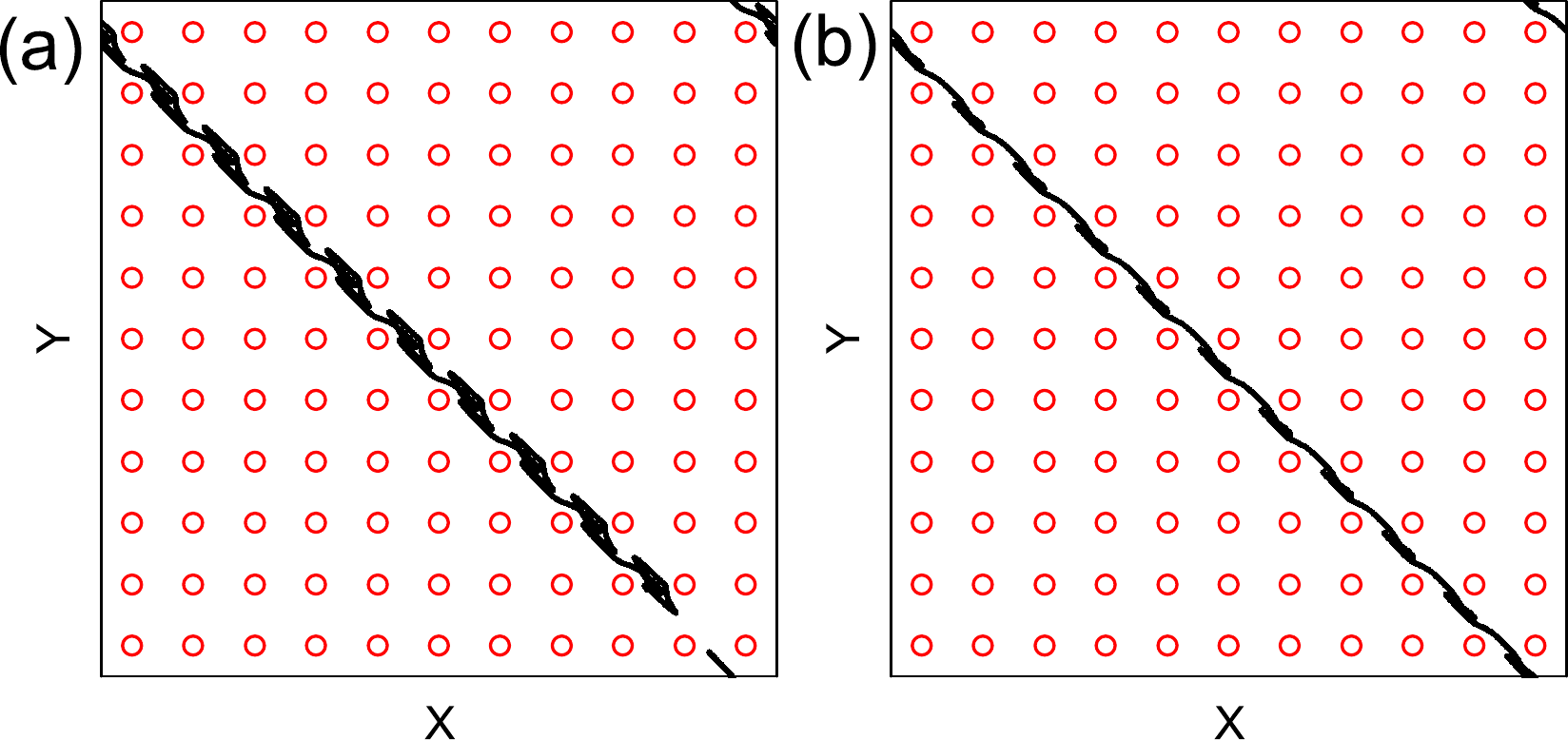}
\caption{ 
Skyrmion trajectory (black line) and obstacle locations (red circles)
for the system in Fig.~\ref{fig:7} with $\alpha_m/\alpha_d=1.0$
and
$x$ direction ac driving of magnitude $A=0.5$.
(a) $F_{D} = 0.1$, where the trajectories are partially disordered.
(b) Along a step at $F_{D} = 0.2$, where the trajectories are more ordered. 
}
\label{fig:8}
\end{figure}

In Fig.~\ref{fig:7}(a) we plot $\langle V_{||}\rangle$ and $\langle V_{\perp}\rangle$
versus $F_{D}$ for a system with
$x$ direction ac driving of
magnitude $A=0.5$ as in Fig.~\ref{fig:3} but with $\alpha_{m}/\alpha_{d} =1.0$, 
where the intrinsic Hall angle is $\theta_{sk}^{int}=45^\circ$.
Figure~\ref{fig:7}(b) shows the corresponding measured
$\theta_{sk}$ which
has only two values, with
$\theta_{sk} = 0^\circ$ at small drives followed by
a jump to the intrinsic value  $\theta_{sk} = -45^\circ$, indicating that there are 
no intermediate directional locking phases. 
Once the system is locked to $-45^\circ$,
a series of Shapiro steps still appear in both the parallel and
perpendicular velocities in Fig.~\ref{fig:7}(a) that 
do not correspond to changes in $\theta_{sk}$.
This shows that it is possible
for Shapiro steps to occur even when
the system motion is fixed along a locking angle.
On the Shapiro steps in Fig.~\ref{fig:7},
the skyrmion trajectories are much more ordered,
as shown in Fig.~\ref{fig:8}(b) at $F_{D} = 0.2$,
while for $F_{D} = 0.1$ in Fig.~\ref{fig:8}(a), the trajectories are less ordered.      
In general, we find that if the ratio
$\alpha_{m}/\alpha_{d}$ produces an intrinsic skyrmion Hall angle that
gives a ratio of $y$ to $x$ motion that is close to
$1/4, 1/3, 1/2$, or $1$, which 
correspond to strong symmetry directions of the substrate lattice,
the system locks permanently to this symmetry direction even for very low drives,
and steps in the velocity appear that are a signature of Shapiro steps
instead of directional locking steps.

\begin{figure}
\includegraphics[width=3.5in]{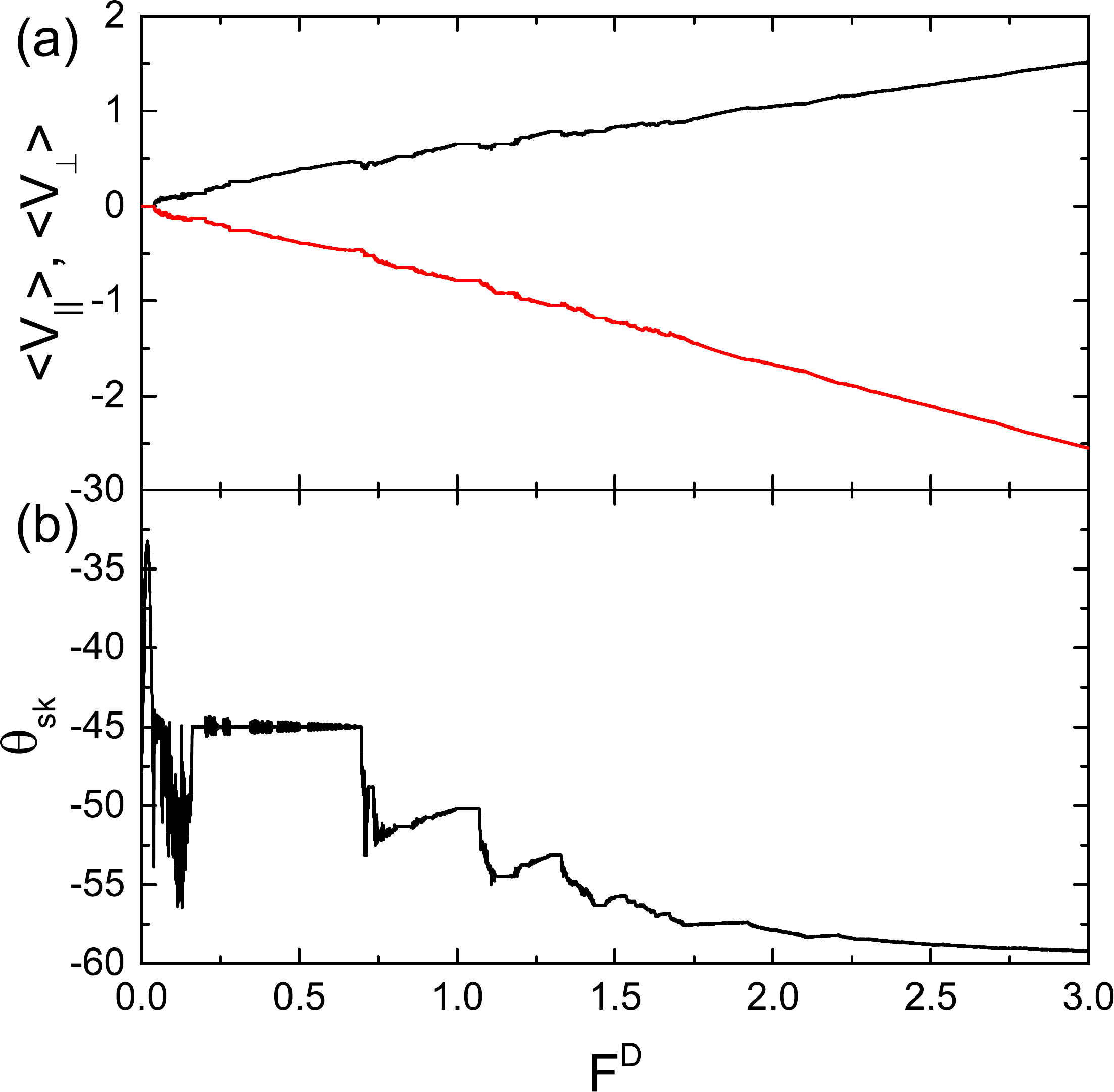}
\caption{ 
(a) $\langle V_{||}\rangle$ (black) and $\langle V_{\perp}\rangle$ (red)
  vs $F_{D}$ for a system with
  $x$ direction ac driving of magnitude
  $A=0.5$ at
  $\alpha_{m}/\alpha_{d} =1.732$.
(b) The corresponding $\theta_{sk}$ vs $F^{D}$. 
}
\label{fig:9}
\end{figure}

\begin{figure}
\includegraphics[width=3.5in]{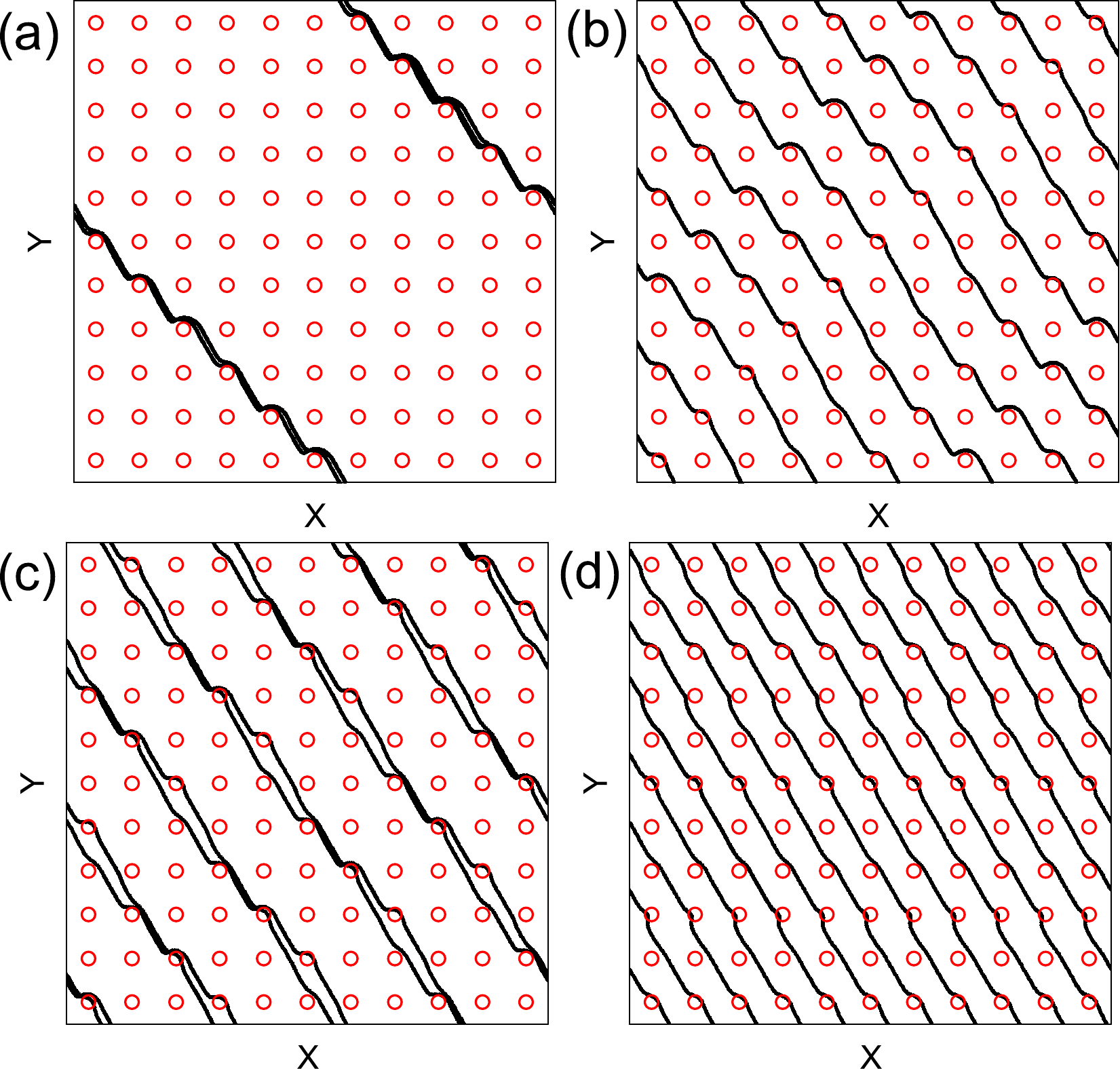}
\caption{ 
Skyrmion trajectory (black line) and obstacle locations (red circles)
for the system in Fig.~\ref{fig:9} with $\alpha_m/\alpha_d=1.732$
and $x$ direction ac driving of magnitude $A=0.5$.
(a) $F^{D} = 0.5$, where the system is locked to $\theta_{sk} = -45^\circ$. 
(b) $F^{D} = 1.0$.
(c) $F^{D}= 1.3$.
(d) $F^{D} = 1.75$ with $\theta_{sk} = -57.5^\circ$.  
}
\label{fig:10}
\end{figure}

In Fig.~\ref{fig:9}(a) we plot $\langle V_{||}\rangle$ and $\langle V_{\perp}\rangle$
versus $F^{D}$ for a system with
$x$ direction ac driving of
magnitude $A=0.5$ as in Fig.~\ref{fig:3} but with
$\alpha_{m}/\alpha_{d} = 1.732$,
giving an intrinsic Hall angle of $\theta_{sk} = 60^{\circ}$.
Figure~\ref{fig:9}(b) shows the corresponding
$\theta_{sk}$ versus $F_{D}$.
Here the system is directionally locked to $\theta_{sk} = 45^\circ$, but there
is still a series of steps in the velocities at low $F^D$ despite the fact that the Hall angle is
constant in this regime.
For $F^{D} > 0.6$, a series
of steps appear in $\theta_{sk}$ as the system switches between
different locking steps.
The larger increases in the magnitude of $\theta_{sk}$ are followed by regions
in which the magnitude of $\theta_{sk}$ decreases by a smaller amount,
and at large $F^{D}$, $\theta_{sk}$ gradually approaches the intrinsic value. 
In Fig.~\ref{fig:10}(a) we illustrate the skyrmion trajectory
for the system in Fig.~\ref{fig:9} at $F^{D} = 0.5$ where the skyrmion
is in the $\theta_{sk} = -45^\circ$ directional locking regime,
while in Fig.~\ref{fig:10}(b) we show the $F^D=1.0$ state
where the skyrmion is locked to an angle close to
$\theta_{sk} = -50^\circ$.
At $F^D=1.3$ in Fig.~\ref{fig:10}(c),
the skyrmion is moving in an alternating fashion.
In Fig.~\ref{fig:10}(d) at $F_{D} = 1.75$,
$\theta_{sk} = -57.5^\circ$.

\begin{figure}
\includegraphics[width=3.5in]{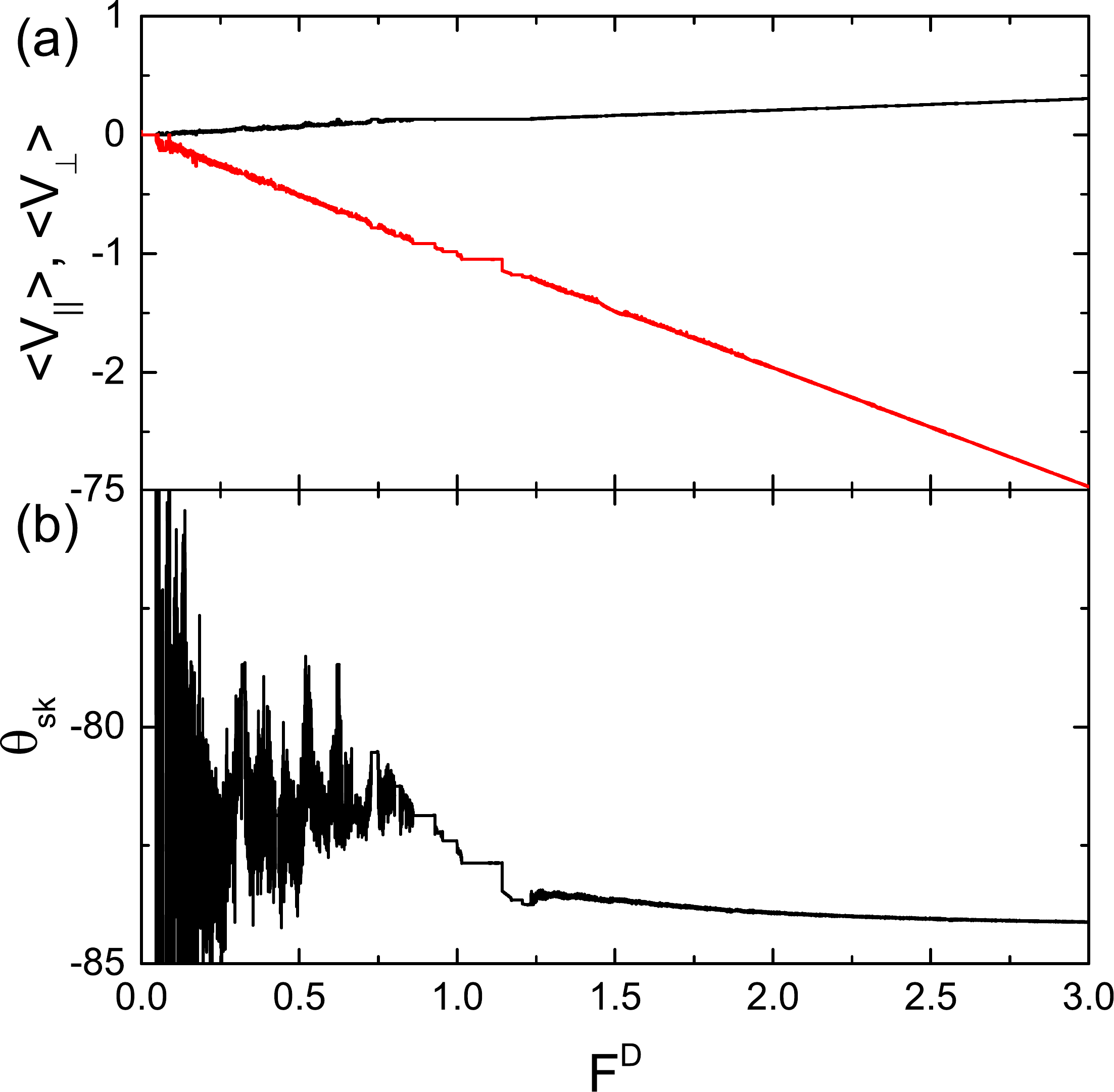}
\caption{ 
(a) $\langle V_{||}\rangle$ (black) and $\langle V_{\perp}\rangle$ (red) vs $F_{D}$
  for a system with
  $x$ direction ac driving of magnitude $A=0.5$
  at $\alpha_{m}/\alpha_{d} =9.962$.
(b) The corresponding $\theta_{sk}$ vs $F^{D}$.
}
\label{fig:11}
\end{figure}

\begin{figure}
\includegraphics[width=3.5in]{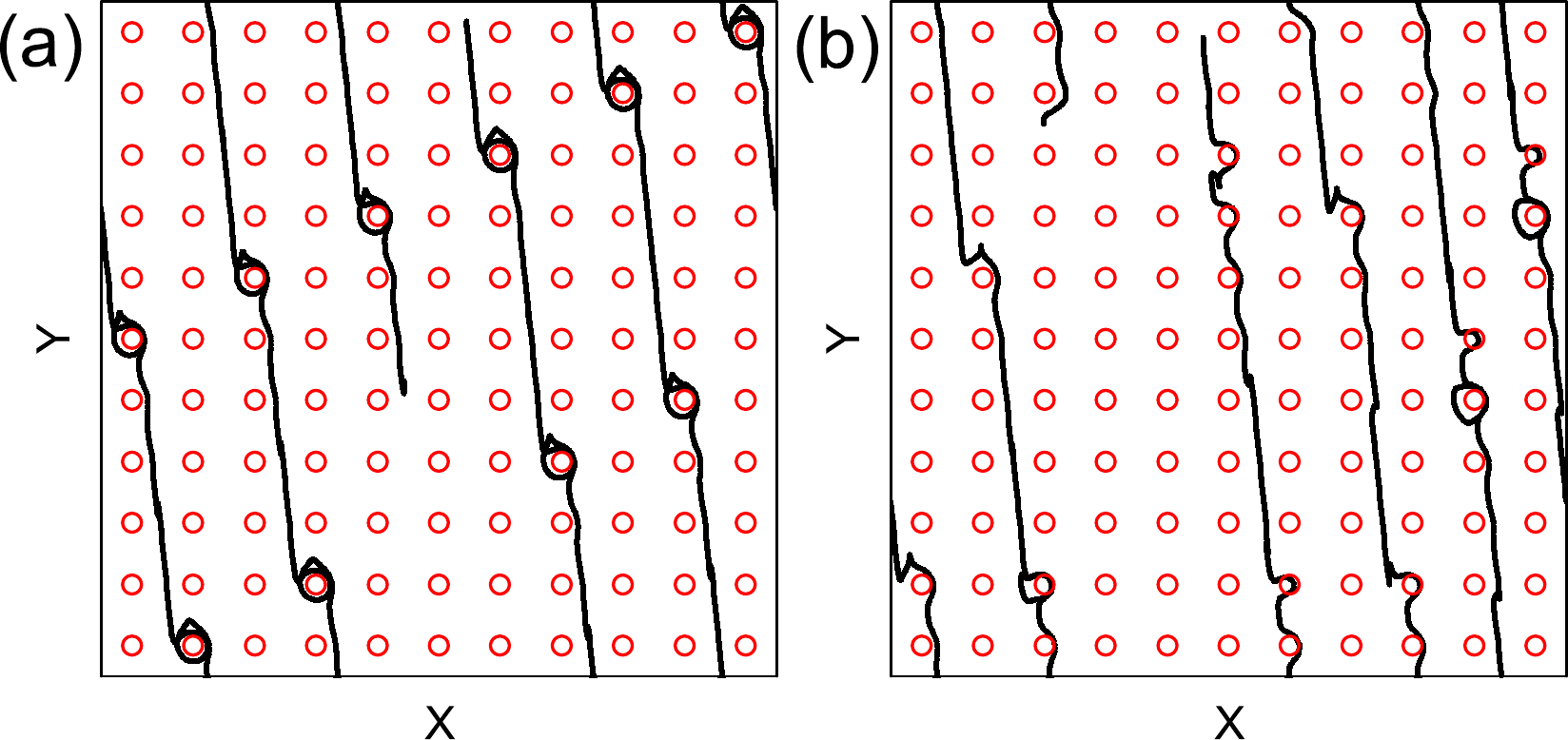}
\caption{ 
  Skyrmion trajectory (black line) and obstacle locations (red circles)
  for the system in Fig.~\ref{fig:12} with $\alpha_m/\alpha_d=9.962$,
  $x$ direction ac driving, and $A=0.5$.
  (a) At $F_{D} = 0.33$, loop orbits appear.
(b) At $F^{D} = 0.5$, the orbits are disordered.
}
\label{fig:12}
\end{figure}

For increasing Magnus force, the dynamics become increasingly disordered,
weakening both the directional locking and the Shapiro steps.
Figure~\ref{fig:11}(a) shows $\langle V_{||}\rangle$ and $\langle V_{\perp}\rangle$
versus $F^{D}$ for a system with
$x$ direction ac driving at $A=0.5$,
as in Fig.~\ref{fig:9}, but for $\alpha_{m}/\alpha_{d} = 9.962$,
where the intrinsic Hall angle is $\theta_{sk} = 84.3^{\circ}$.
Figure~\ref{fig:11}(b) shows the corresponding
$\theta_{sk}$ versus $F_{D}$.
In this case there are only small steps in the velocity force curves 
that are associated with steps in $\theta_{sk}$,
which has an average value near $\theta_{sk}=-82.5^\circ$.
For these higher Magnus forces,
the skyrmion starts to perform full or partial loops around
the obstacles, as shown in 
Fig.~\ref{fig:12}(a) at $F_{D} = 0.33$.
In Fig.~\ref{fig:12}(b) at $F^{D} = 0.5$, the system is in a disordered phase.
For $F_{D} > 1.5$,
the locking regimes are lost and
$\theta_{sk}$ gradually approaches the intrinsic Hall angle value.  

\section{AC Driving in the Transverse Direction}

\begin{figure}
\includegraphics[width=3.5in]{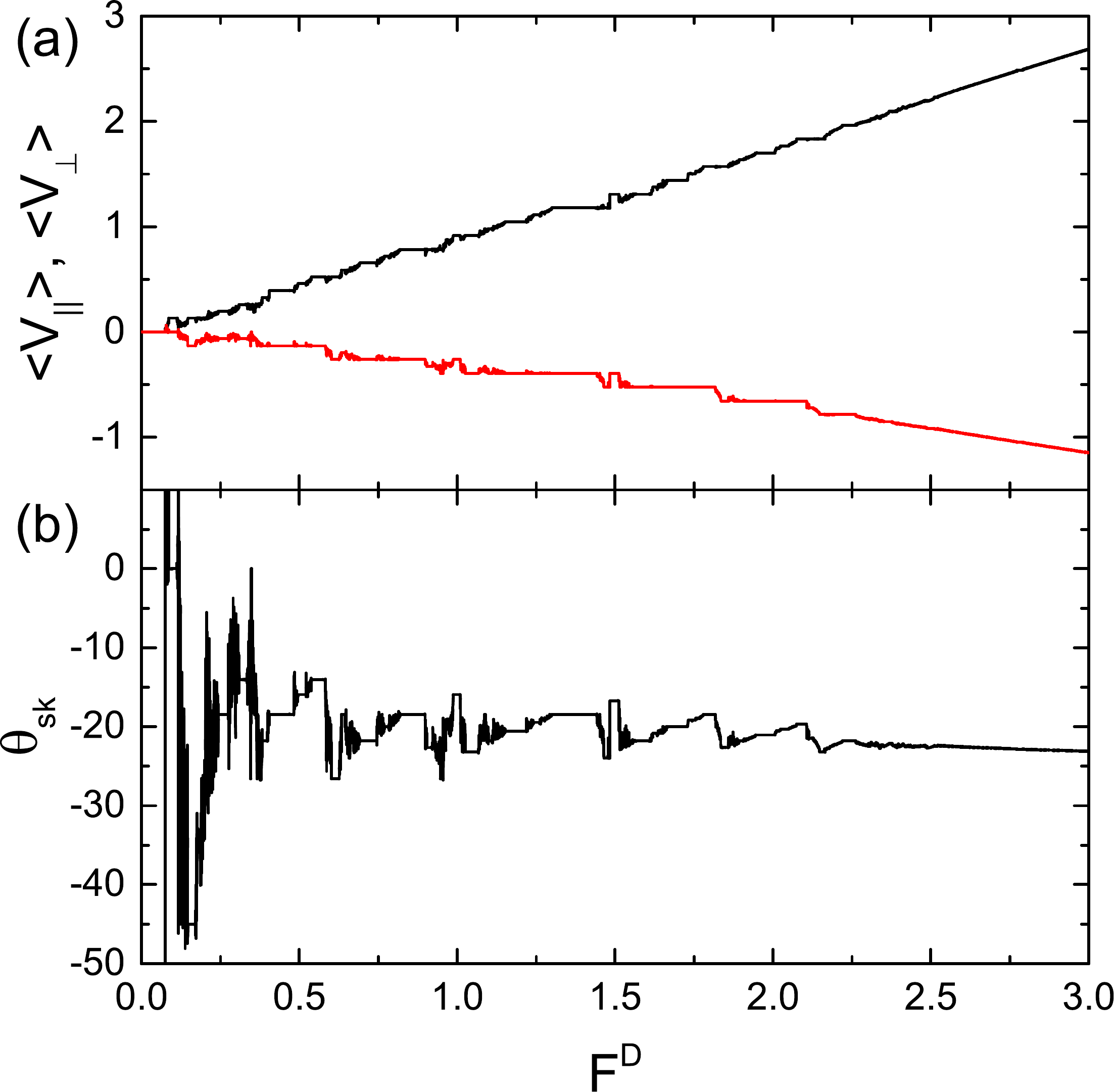}
\caption{ (a) $\langle V_{||}\rangle$ (black) and $\langle V_{\perp}\rangle$ (red)
  for a system with $\alpha_m/\alpha_d=0.45$
  and $y$ direction
  ac driving of magnitude $A=0.5$.
  (b) The corresponding $\theta_{sk}$ vs $F^{D}$.
  We find more steps than for the same system with
  ac driving in the $x$ direction (Fig.~\ref{fig:3}).  
}
\label{fig:13}
\end{figure}

\begin{figure}
\includegraphics[width=3.5in]{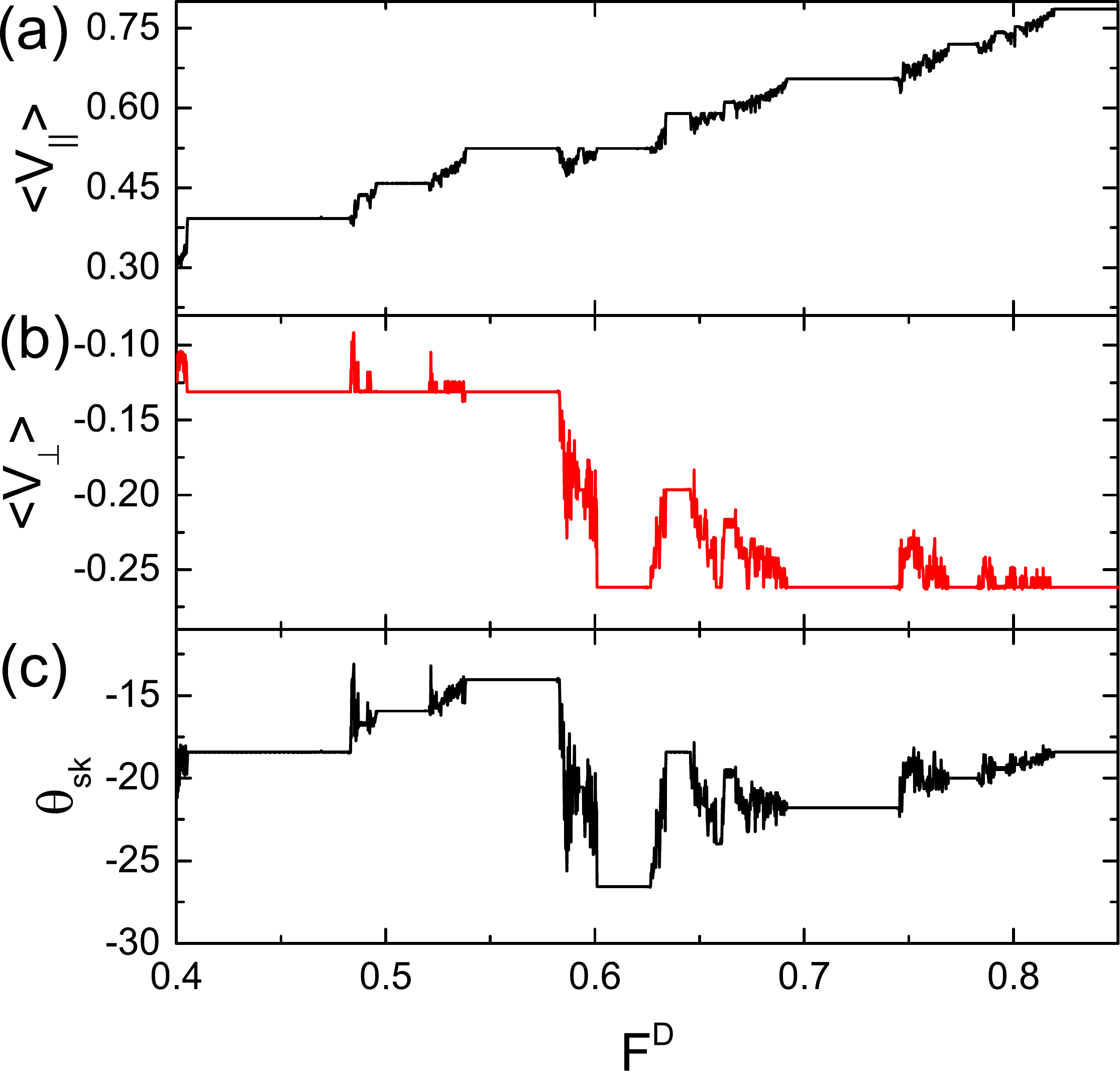}
\caption{(a) $\langle V_{||}\rangle$,
  (b) $\langle V_{\perp}\rangle $, and (c)
  $\theta_{sk}$ vs $F_{D}$
  for the system in Fig.~\ref{fig:13} with $\alpha_m/\alpha_d=0.45$,
  $A=0.5$, and $y$ direction ac driving shown over
the interval $0.4 < F^{D} < 0.85$.}
\label{fig:14}
\end{figure}

\begin{figure}
\includegraphics[width=3.5in]{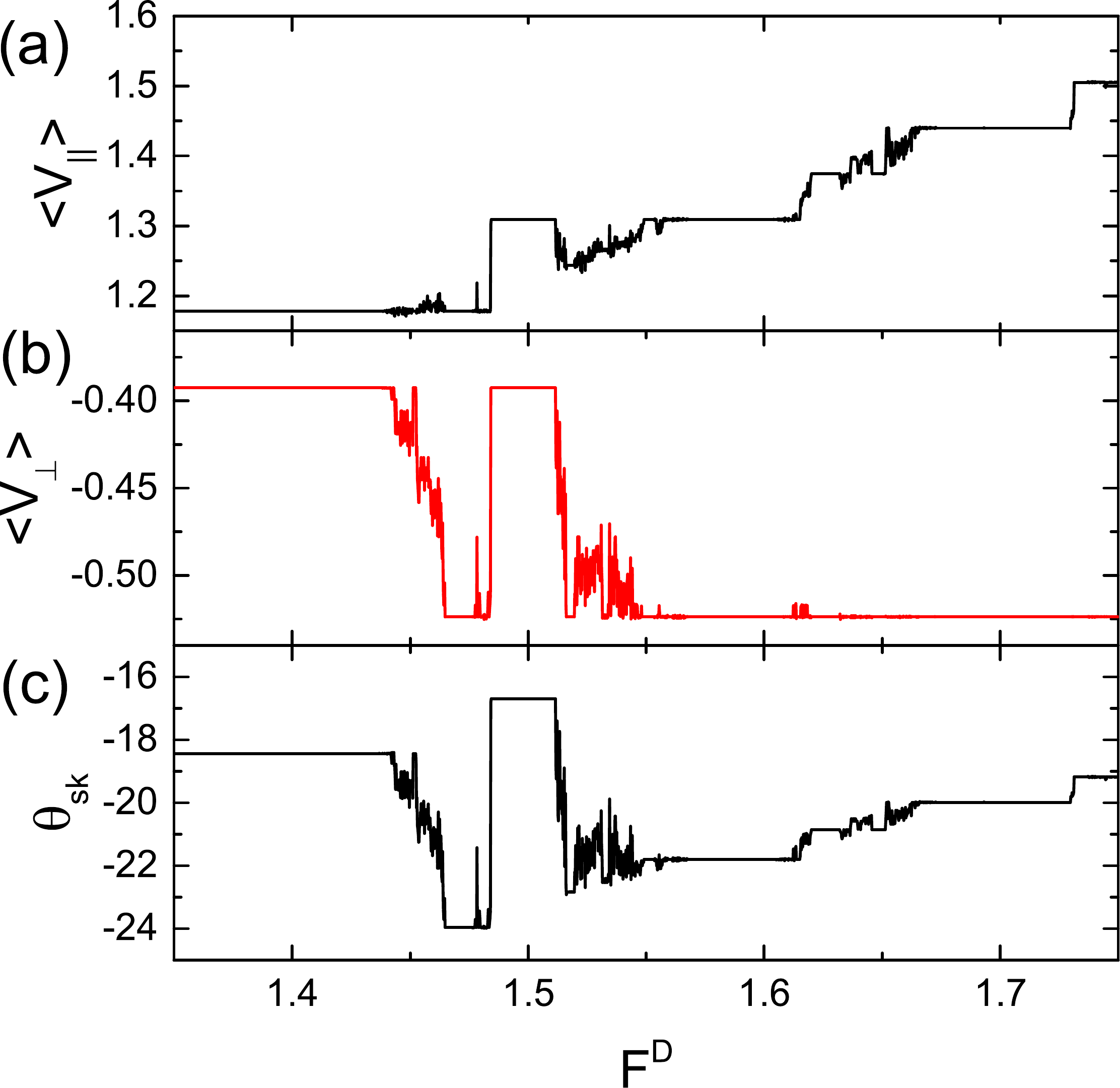}
\caption{ 
(a) $\langle V_{||}\rangle$,
  (b) $\langle V_{\perp}\rangle $, and (c)
$\theta_{sk}$ vs $F_{D}$ for the system in Fig.~\ref{fig:13}
  with $\alpha_m/\alpha_d=0.45$,
  $A=0.5$, and $y$ direction ac driving shown over
  the interval $1.35 < F^{D} < 1.75$.
}
\label{fig:15}
\end{figure}

We next consider the case where the
ac drive is applied along the $y$-direction, transverse to the dc drive. 
In Fig.~\ref{fig:13}(a) we plot the velocity components versus
$F^{D}$ and in Fig.~\ref{fig:13}(b) we show the corresponding $\theta_{sk}$
versus $F^{D}$ for a 
system with $\alpha_{m}/\alpha_{d} = 0.45$
and $A=0.5$.
The features in the velocity curves are more step like, rather than the
cusp like shapes found for $x$ direction ac driving in Fig.~\ref{fig:3},
and in general there are more locking regions which are associated
with both directional locking
and the ac phase locking.
Another interesting feature is that near $F^{D}=0.15$, there is a window
of locking to
$\theta_{sk} = -45^\circ$,
which is considerably larger in magnitude
than the intrinsic skyrmion Hall angle 
of $\theta_{sk} =24.2^\circ$.  We call this a Hall angle overshoot.
As $F_{D}$ increases, $\theta_{sk}$ undergoes a
number of oscillations until it reaches 
a saturation near the intrinsic value at high $F^{D}$.   
In Fig.~\ref{fig:14}
we plot $\langle V_{||}\rangle$, $\langle V_{\perp}\rangle$,
and $\theta_{sk}$ versus $F^{D}$ for the system in
Fig.~\ref{fig:13}
over the interval $0.4 < F^{D} < 0.85$.
There are sudden jumps both up and down in $\theta_{sk}$.
Additionally, there
are regions where $\langle V_{\perp}\rangle$ remains
constant but steps appear in $\langle V_{||}\rangle$
that are associated with
jumps in $\theta_{sk}$. 
In the interval $1.35 < F^D <1.75$ shown
in Fig.~\ref{fig:15}(a,b,c),
there are regions where the velocity can decrease with increasing $F_D$.

\begin{figure}
\includegraphics[width=3.5in]{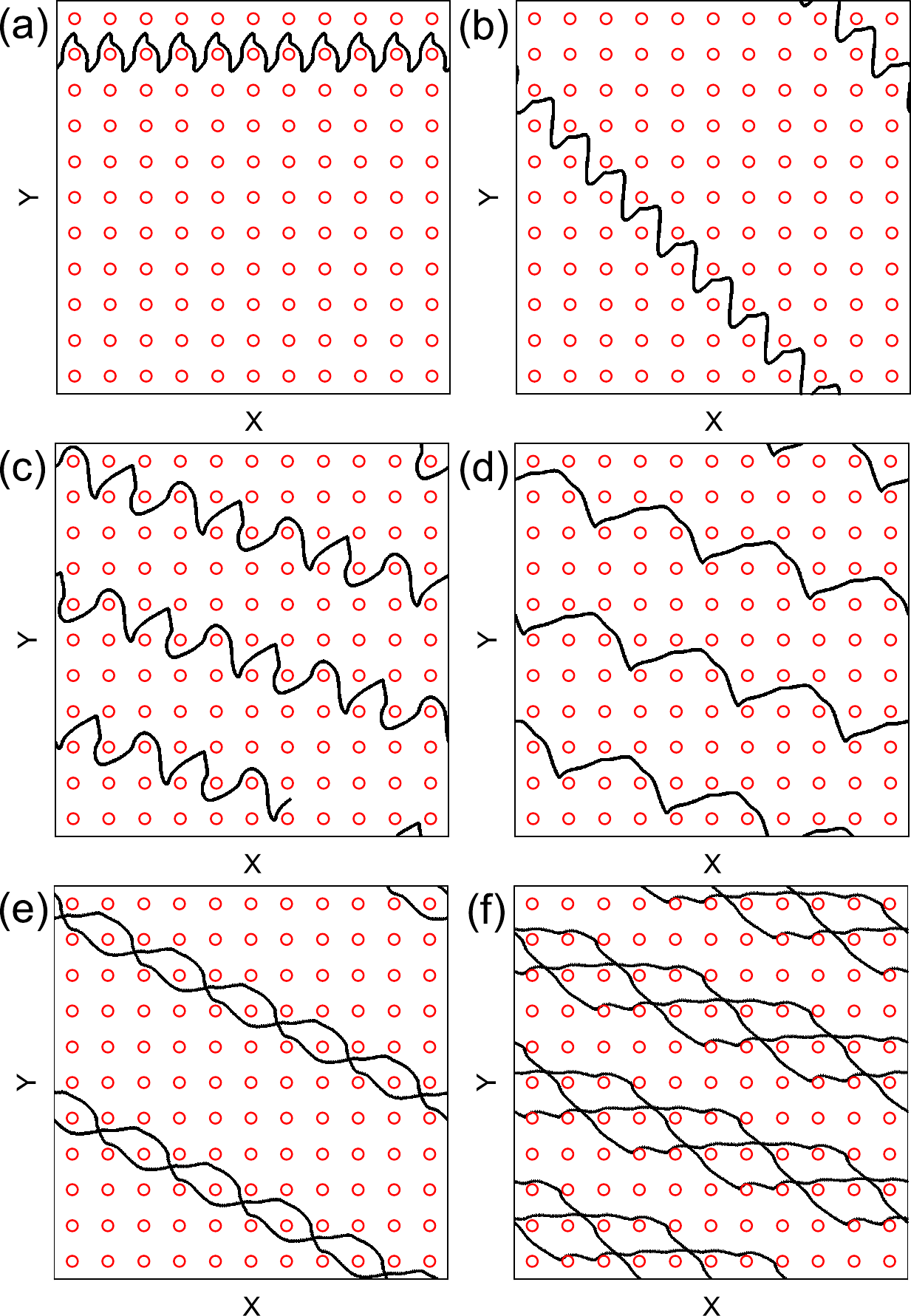}
\caption{ 
Skyrmion trajectory (black line) and obstacle locations (red circles)
for the system in Fig.~\ref{fig:13} with $\alpha_m/\alpha_d=0.45$,
$A=0.5$,
and $y$ direction ac driving.
(a) At $F^{D} = 0.1$ the motion is aligned with the $x$ direction. 
(b) At $F^{D} = 0.16$, the motion is locked to $\theta_{sk} = -45^\circ$.
(c) At $F^{D} = 0.26$, the motion is at a smaller angle of $\theta_{sk} = -18.4^\circ$.
(d) $F^{D} = 0.44$.  (e) $F^{D} = 0.61$. (f)$ F^{D} = 1.4$.  
}
\label{fig:16}
\end{figure}

In Fig.~\ref{fig:16} we show some of the representative skyrmion
orbits for the system in Fig.~\ref{fig:13}.
At $F^D=0.1$ in Fig.~\ref{fig:16}(a),
the motion is locked in the $x$ direction, and the skyrmion executes a zig-zag pattern. 
In Fig.~\ref{fig:16}(b) at $F^{D} = 0.16$,
the motion is locked to $\theta_{sk} = -45^\circ$.
Figure~\ref{fig:16}(c) shows the
trajectory at $F^{D} = 0.26$,
where the skyrmion moves at a much smaller angle of
$\theta_{sk} \approx -18.4^{\circ}$.
In Fig.~\ref{fig:16}(d) at $F^{D} = 0.44$,
we still find $\theta_{sk} = 18.4^\circ$ but the orbit
shape has changed, with
the skyrmion moving $3a$ in
the $x$ direction and $a$ in the $y$ direction during each ac drive cycle.
At $F^D=0.61$ in 
Fig.~\ref{fig:16}(e),
the motion is along
$\theta_{sk} = -26.6^\circ$, and in Fig.~\ref{fig:16}(f)
at $F^{D} = 1.4$,
$\theta_{sk} = -18.4^\circ$,
indicating that the system has returned to the $1/3$ locking region. 
The orbit differs 
from that shown in Fig.~\ref{fig:16}(d),
indicating that $R = 1/3$ locking can occur in several different ways.

\begin{figure}
\includegraphics[width=3.5in]{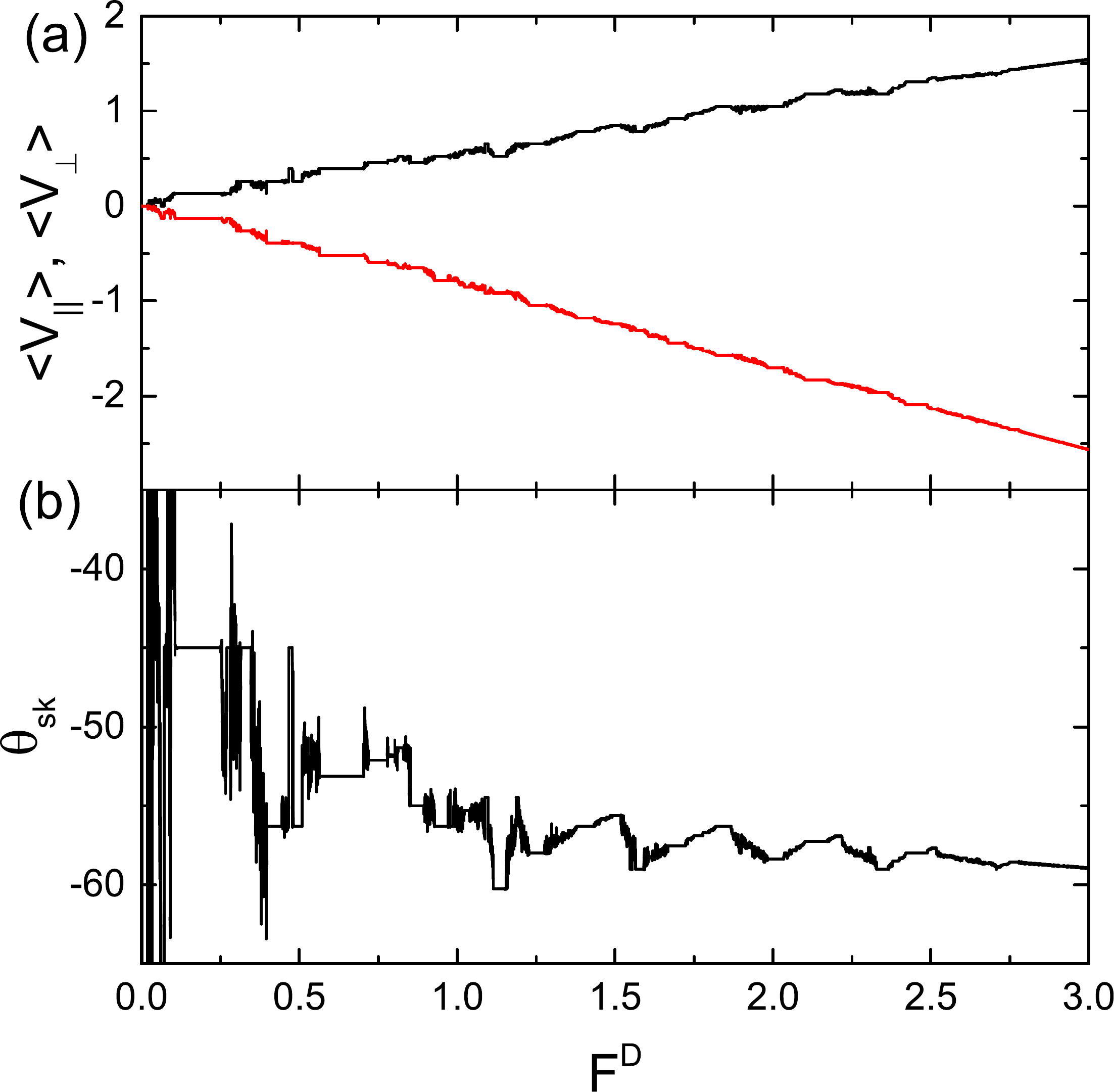}
\caption{ 
  (a) $\langle V_{||}\rangle$ (black) and $\langle V_{\perp}\rangle$ (red)
  vs $F^{D}$ for
  a system with $\alpha_{m}/\alpha_{d} = 1.732$,
  $A=0.5$, and $y$ direction ac driving.
  (b) The corresponding $\theta_{sk}$ vs $F^{D}$.   
}
\label{fig:17}
\end{figure}

\begin{figure}
\includegraphics[width=3.5in]{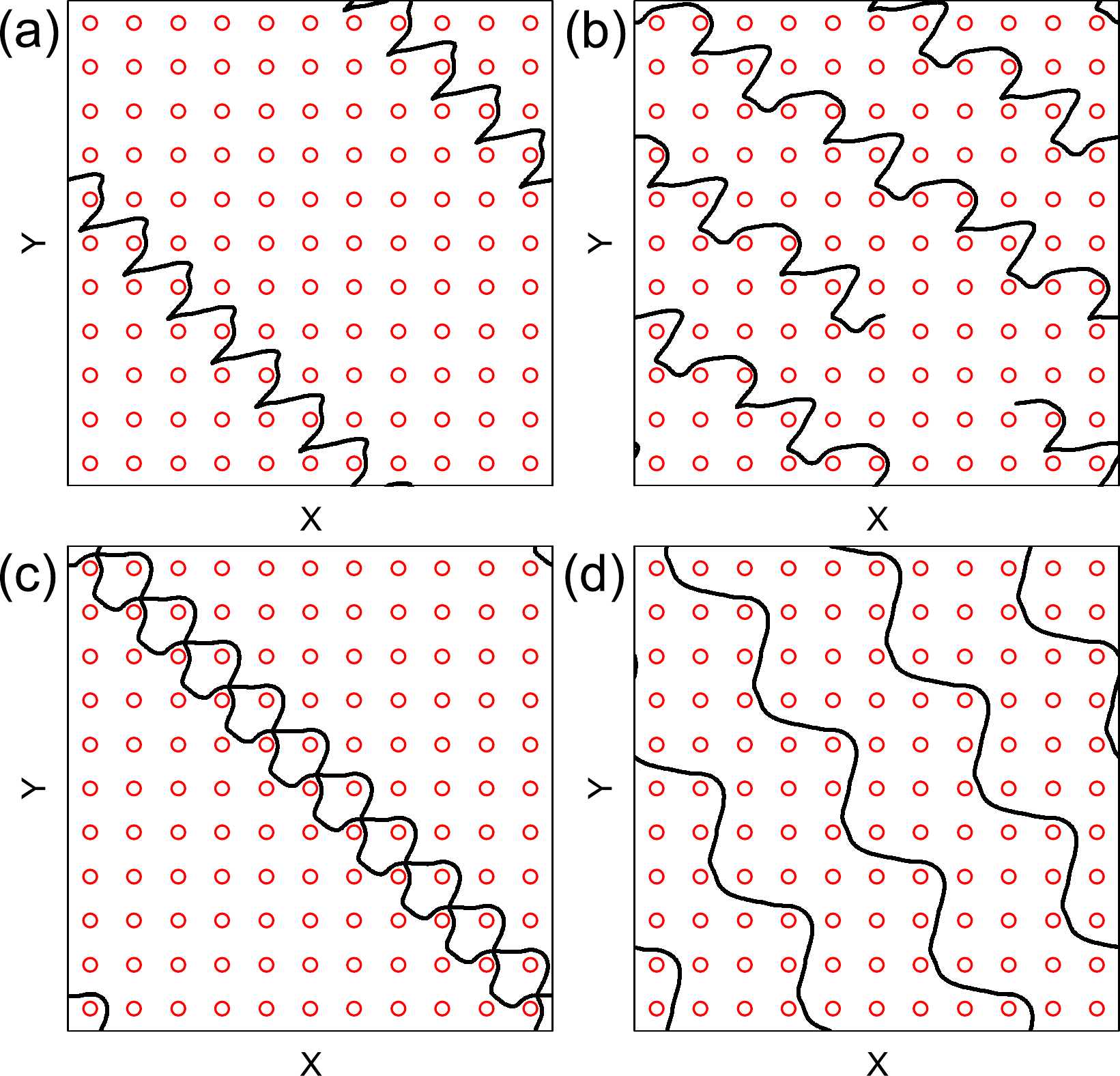}
\caption{ 
  Skyrmion trajectory (black line) and obstacle locations (red circles)
  for the system in Fig.~\ref{fig:17} with $\alpha_m/\alpha_d=1.732$,
  $A=0.5$, and $y$ direction ac driving.
  (a) At $F^{D} = 0.15$, the motion is locked to $\theta_{sk} = -45^\circ$. 
  (b) At $F^{D} = 0.175$, the motion locks to $\theta_{sk} = -33^\circ$.
  (c) At $F^{D} = 0.3$, the motion is along $\theta_{sk} = -45^\circ$.
  (d) At $F^{D} = 0.43$, the motion locks to $\theta_{sk} = -56.3^\circ$.  
}
\label{fig:18}
\end{figure}

In Fig.~\ref{fig:17}(a) we plot the velocity curves versus $F^{D}$
and in Fig~\ref{fig:17}(b) we show the corresponding
$\theta_{sk}$ versus $F^{D}$ for a system with $\alpha_{m}/\alpha_{d} = 1.732$,
where there are again a series of steps at which
$\theta_{sk}$ increases or decreases.
Locking occurs in several regimes and
the system jumps in and out of the
$\theta_{sk} = -45^\circ$ locked state since
the $45^\circ$ locking is a particularly strong symmetry direction of
the square obstacle lattice.  
In Fig.~\ref{fig:18}(a,b) we plot the skyrmion trajectories 
for the system in Fig.~\ref{fig:17} at
$F^{D} = 0.15$ in the $-45^{\circ}$ locking regime
and at $F^{D} = 0.185$, where the skyrmions move at a
lower magnitude angle
of $\theta_{sk} = -33.7^\circ$.
In Fig.~\ref{fig:18}(c) at $F^{D}  = 0.3$,
the system jumps to a new $-45^\circ$ locking phase with
a braiding patten, and in Fig.~\ref{fig:18}(d)
at $F_{D} =0.43$, the motion is along $\theta_{sk}  = -56.3^\circ$.

\begin{figure}
\includegraphics[width=3.5in]{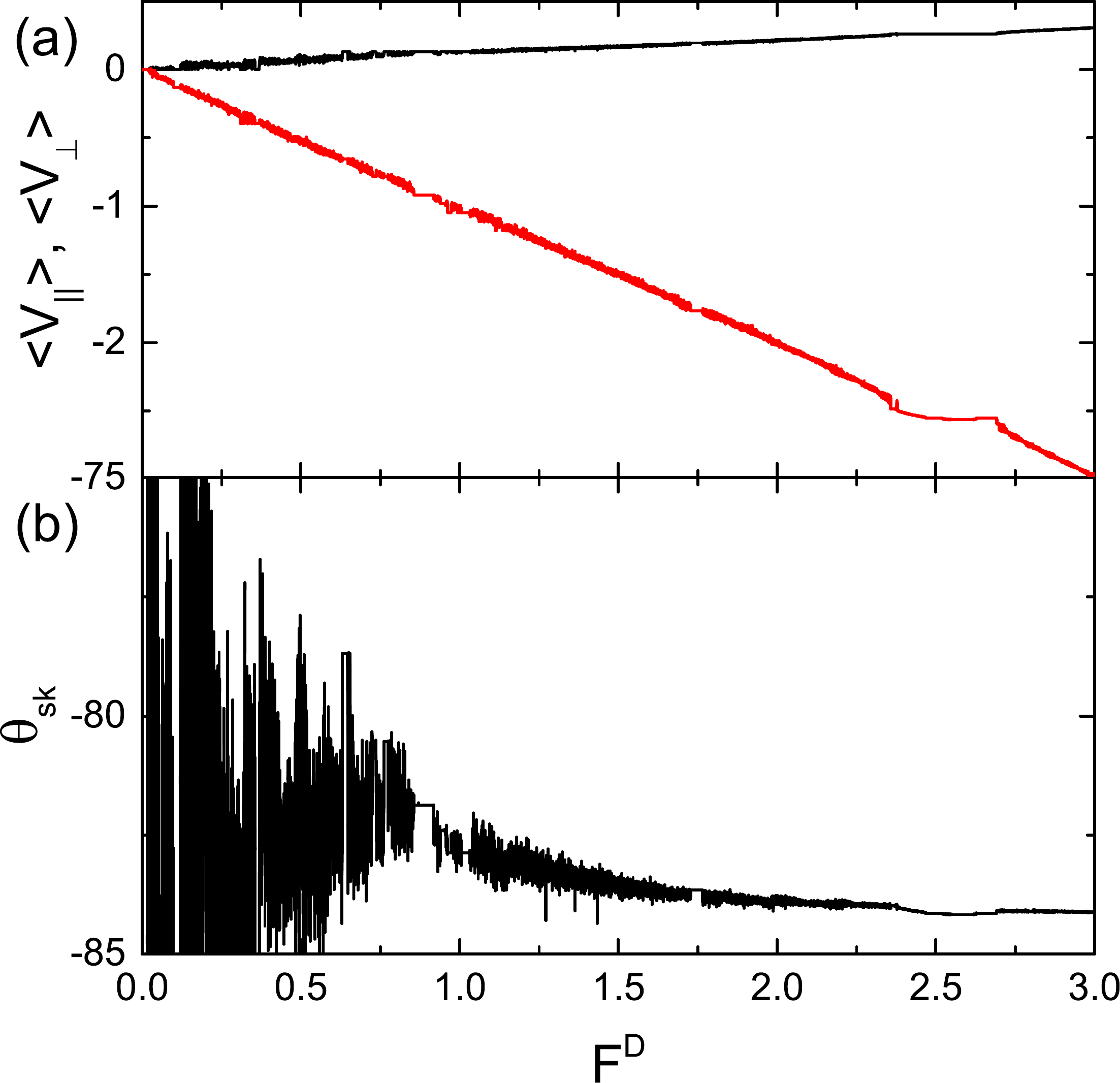}
\caption{ 
  (a) $\langle V_{||}\rangle$ (black) and  $\langle V_{\perp}\rangle$ (red)
  vs $F^D$
  for
  a system with $\alpha_{m}/\alpha_{d} = 9.962$,
  $A=0.5$, and $y$ direction ac driving.
  (b) The corresponding $\theta_{sk}$ vs $F^{D}$.
}
\label{fig:19}
\end{figure}

\begin{figure}
\includegraphics[width=3.5in]{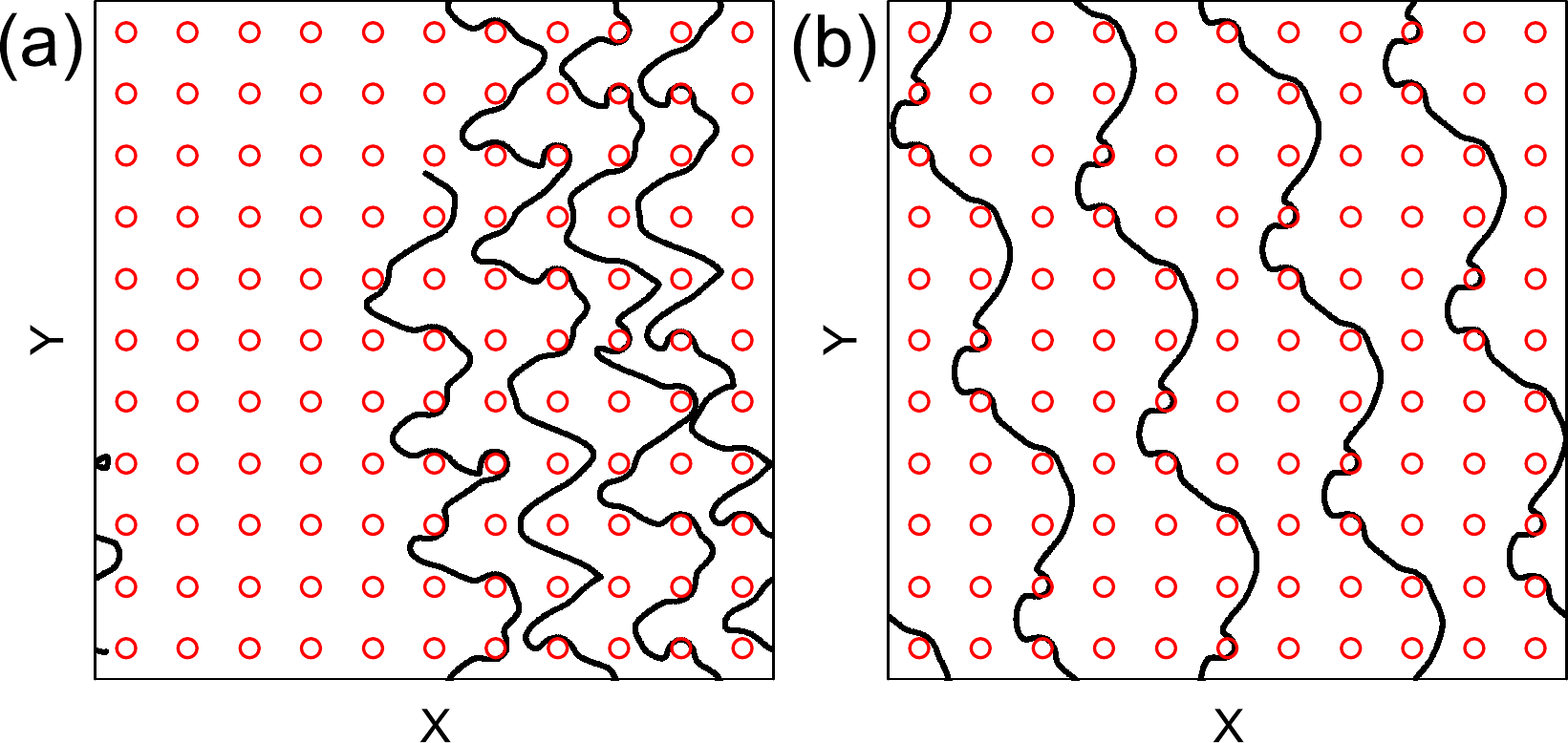}
\caption{ 
  Skyrmion trajectory (black line) and obstacle locations (red circles)
  for the system in Fig.~\ref{fig:19} with $\alpha_m/\alpha_d=9.962$,
  $A=0.5$, and $y$ direction ac driving.
  (a) $F^{D} = 0.25$ in a non-phase locked region. 
(b) $F^{D} = 0.465$ in a phase locked region with $\theta_{sk} = -76^\circ$.  
}
\label{fig:20}
\end{figure}

For higher values of  $\alpha_{m}/\alpha_{d}$, we again observe
extended regions in which the trajectories are disordered,
and the phase locking phenomena is generally reduced. 
In Fig.~\ref{fig:19} we show the velocities and skyrmion Hall angle versus
$F^{D}$ for a system with
$A=0.5$ and $y$ direction ac driving as
in Fig.~\ref{fig:17} but for
$\alpha_{m}/\alpha_{d} = 9.962$.
There are a number of smaller steps,
particularly in the range $0.35 < F^{D} < 1.0$,
along with one larger step near $F^{D} = 2.5$.
Figure~\ref{fig:20}(a) illustrates
the skyrmion trajectories for the system in
Fig.~\ref{fig:19} at $F^{D} = 0.25$, where
there is no phase locking state and
the trajectories form a non-repeating pattern.
In Fig.~\ref{fig:20}(b)
at $F^{D} = 0.465$, 
the system is phase locked at
$\theta_{sk} = -76^\circ$ and the trajectories are ordered.

For ac driving in the $y$-direction,
there is an interplay between three types of phase locking.  These are the
Shapiro steps, 
the directional locking, and the transverse phase locking effect.
This is the reason that
there are
a larger number of steps in the velocity and skyrmion Hall angle curves
compared to ac driving in the $x$ direction.  

\subsection{Hall Angle Reversal}

\begin{figure}
\includegraphics[width=3.5in]{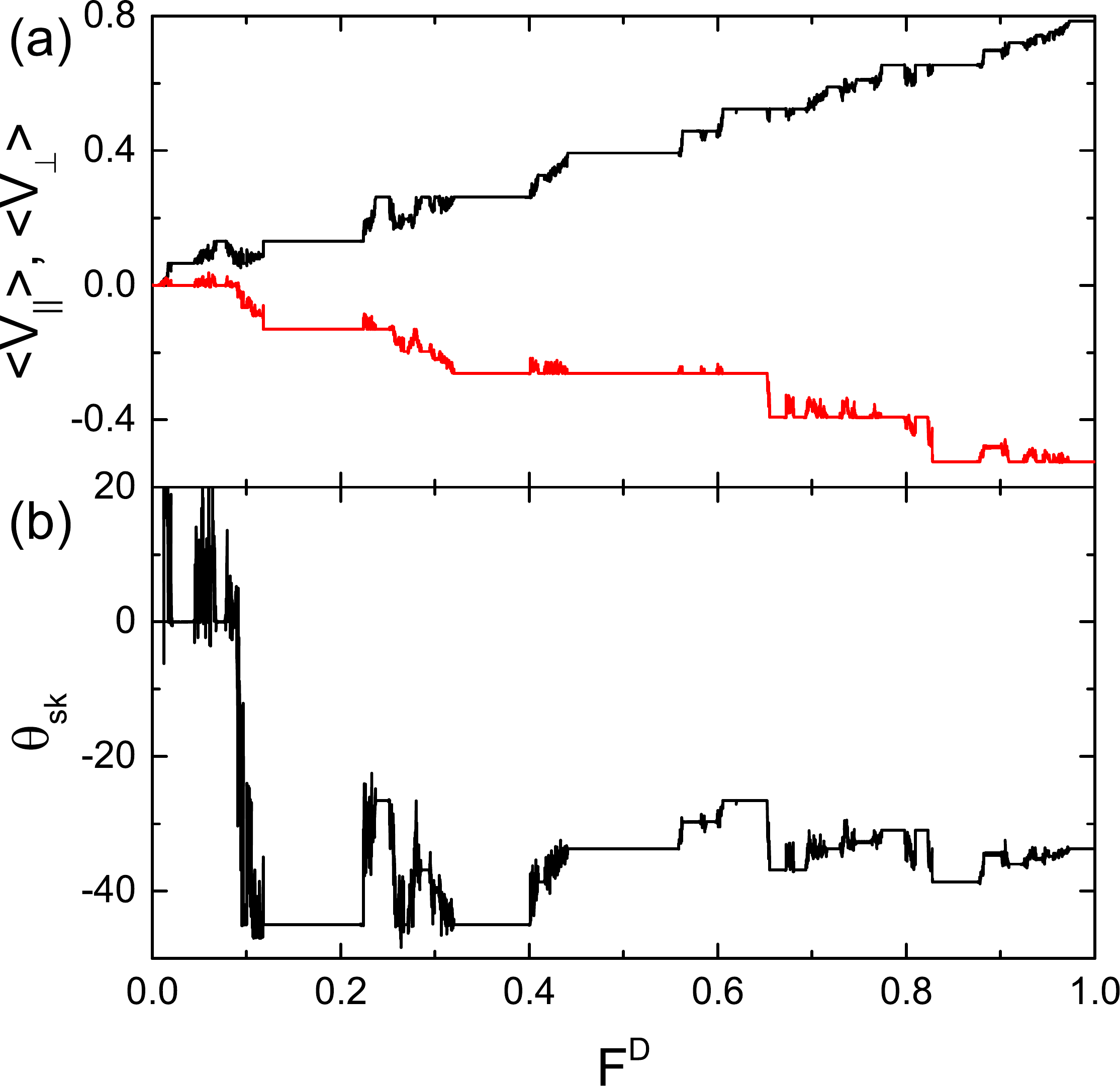}
\caption{(a) $\langle V_{||}\rangle$ (black) and $\langle V_{\perp}\rangle$ (red)
  vs $F_{D}$ for a system with $\alpha_m/\alpha_d=1.0$,
  $A=0.5$, and
  $y$ direction ac driving.
  (b) The corresponding $\theta_{sk}$ vs $F_{D}$.
  For $F_{D} < 0.1$, there are regions
  in which both velocity components are positive, giving a positive skyrmion Hall angle.   
}
\label{fig:21}
\end{figure}

\begin{figure}
\includegraphics[width=3.5in]{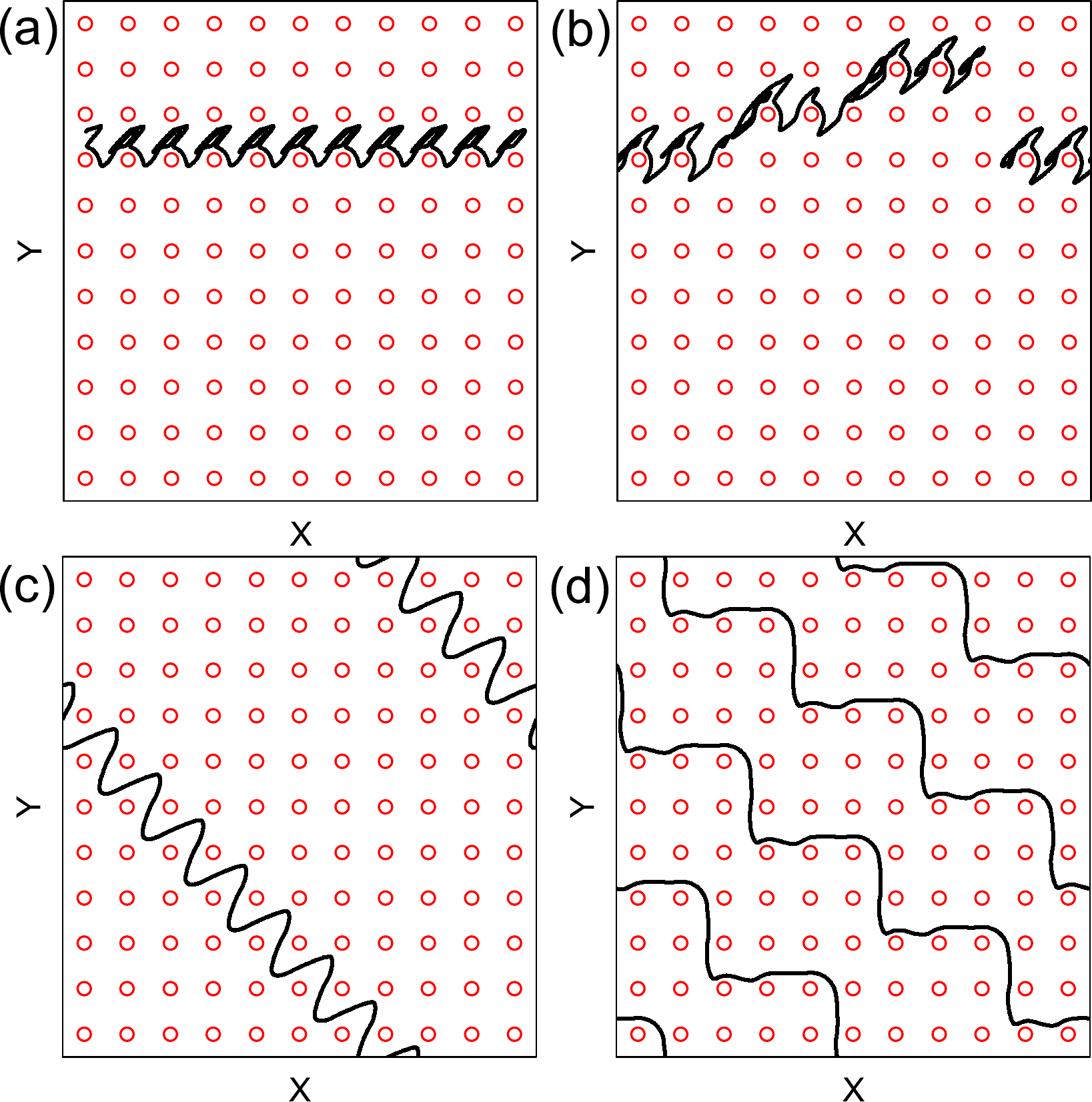}
\caption{Skyrmion trajectory (black line) and obstacle locations (red circles)
  for the system in Fig.~\ref{fig:21} with
  $\alpha_m/\alpha_d=1.0$,
  $A=0.5$, and
  $y$ direction ac driving.
  (a) At $F_{D} = 0.05$, the motion is locked along $x$.
  (b) At $F_{D} = 0.065$, the skyrmion is also translating along the positive $y$
  direction, giving a positive skyrmion Hall angle.
  (c) At $F_{D} = 0.2$, there is locking at
  $\theta_{sk} = -45^\circ$.
  (d) At $F_{D} = 0.55$ there is locking 
at $\theta_{sk} = -33.7^\circ$.  
}
\label{fig:22}
\end{figure}

In most cases, we have shown that although the skyrmion Hall
angle increases or decreases with drive, it maintains the same sign.
Under certain circumstances, however, we find regions in which the
skyrmion Hall angle changes from positive to negative.
This effect is generally associated with windows of disordered motion
at smaller $F^D$
where the skyrmion is jumping among different orbits.
In Fig.~\ref{fig:21} we plot $\langle V_{||}\rangle$, $\langle V_{\perp}\rangle$,
and $\theta_{sk}$ versus $F^D$
for a system with $y$ direction ac driving
at $\alpha_{m}/\alpha_{d} = 1.0$.
If the ac driving were in the $x$-direction, this ratio of the
Magnus to damping terms would produce a constant
skyrmion Hall angle of $\theta_{sk} = 45^\circ$
with only Shapiro steps.
When the ac driving is along the $y$ direction, however,
a variety of locking regions appear that are associated with
jumps both up and down in
$\langle V_{||}\rangle$ and $\langle V_{\perp}\rangle$.
Jumps also occur in
$\theta_{sk}$
among the values $\theta_{sk} = -45^\circ$,
$-38.65^\circ$, $-36.87^\circ$, $-33.6^\circ$, and $-26.56^\circ$. 
The corresponding velocity ratios are
$\langle V_{\perp}\rangle/\langle V_{||}\rangle = 1$, 4/5, 4/3, 2/3, and $1/2$,
respectively.
At
higher drives,
$\theta_{sk}$ decreases in magnitude
to
angles smaller than $45^\circ$.
Meanwhile, for $F_{D} < 0.1$
there
are several regions
in which $\langle V_{||}\rangle$ and $\langle V_{\perp}\rangle$
are both finite but positive,
which produces a positive skyrmion Hall angle of
$\theta_{sk} \approx 10^\circ$.
The motion in this regime is illustrated
in Fig.~\ref{fig:22}(a,b)
at $F_{D} = 0.045$, where the motion is locked along $x$, and
at $F_{D} = 0.065$,
where the skyrmion is jumping intermittently in the positive $y$-direction.
Figure~\ref{fig:22}(c) shows the locking phase with
$\theta_{sk} = -45^\circ$ at $F^{D} = 0.2$, and 
in Fig.~\ref{fig:22}(d) at $F^D=0.55$,
$\theta_{sk} = -33.7^\circ$.
It is possible that by varying other parameters
such as the size of the
obstacles,
clear regions of skyrmion Hall angle reversals
will also emerge, but the results above indicate
that such reversal effects 
can arise for skyrmion motion on periodic substrates. 

\section{Changing AC Amplitude}

\begin{figure}
\includegraphics[width=3.5in]{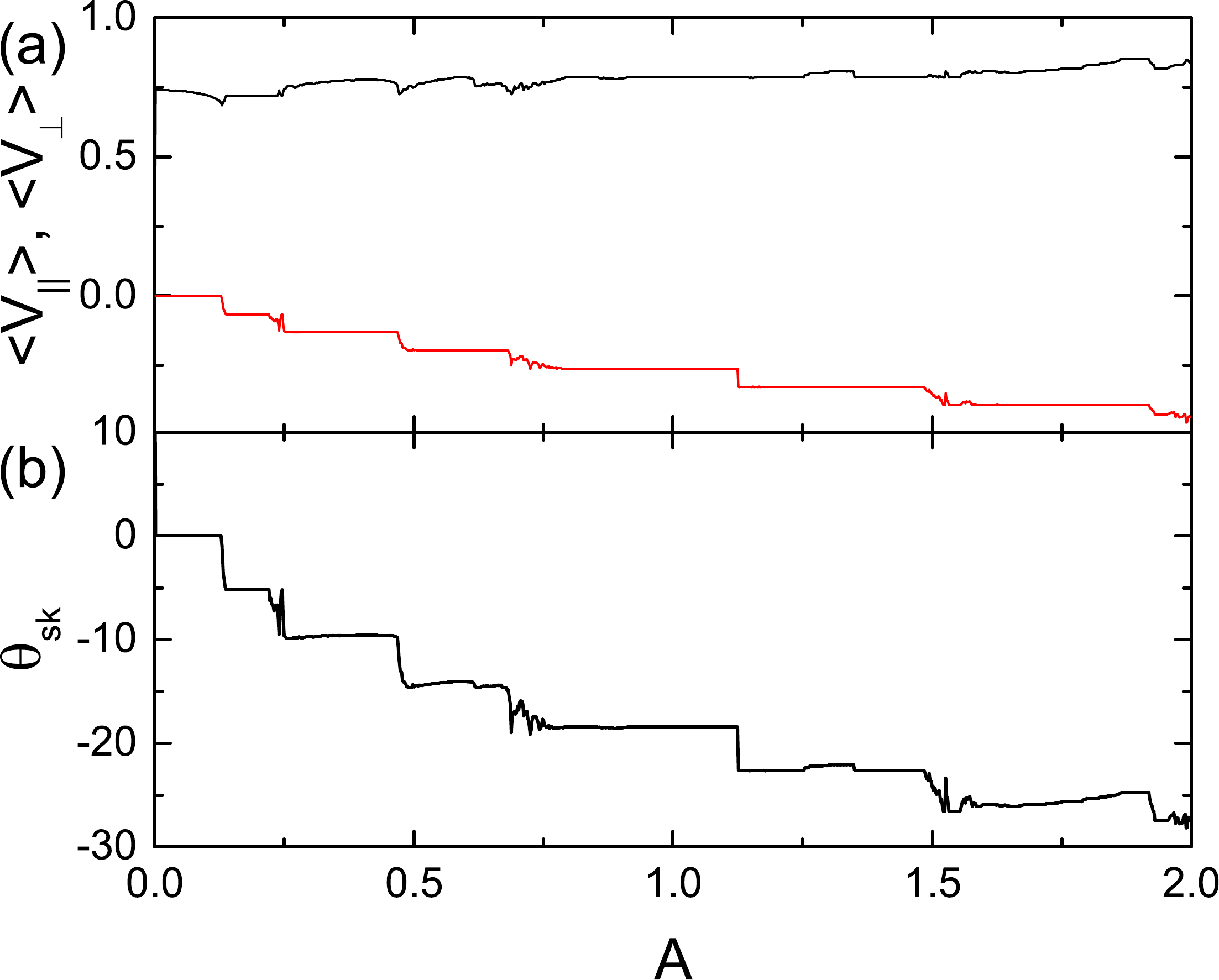}
\caption{ 
  (a) $\langle V_{||}\rangle$ (black) and $\langle V_{\perp}\rangle$ (red) vs ac drive
  amplitude $A$ for a system with $\alpha_m/\alpha_d=0.45$,
  fixed
  $F^D=1.0$, and $x$ direction ac driving.
  (b) The corresponding $\theta_{sk}$ vs $F^{D}$.
  Here the magnitude of the
  skyrmion Hall angle increases in a series of steps with increasing
$A$. 
}
\label{fig:23}
\end{figure}

\begin{figure}
\includegraphics[width=3.5in]{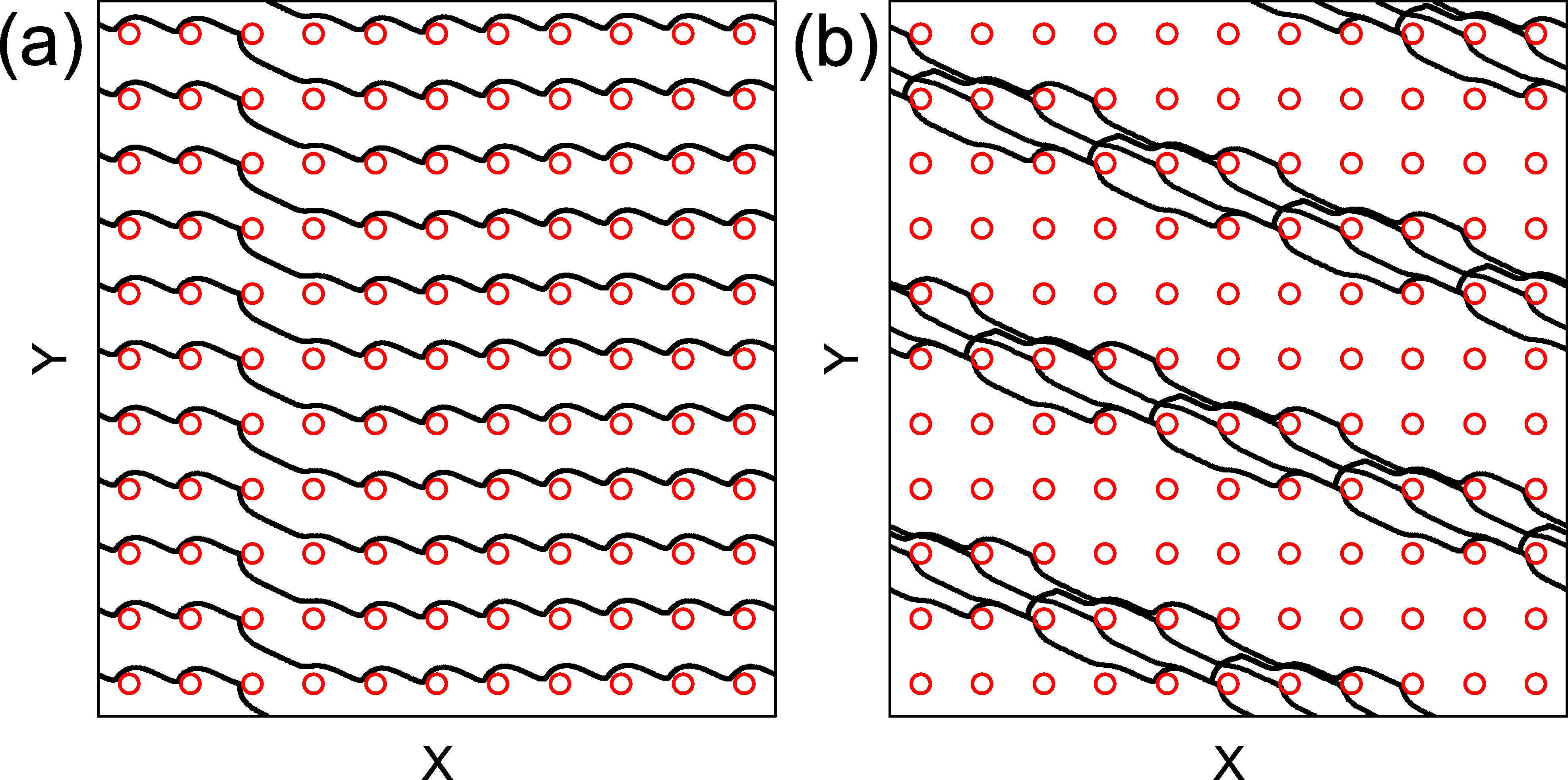}
\caption{ 
  Skyrmion trajectory (black line) and obstacle locations (red circles)
  for the system in Fig.~\ref{fig:23} with $\alpha_m/\alpha_d=1.0$,
  $F^D=1.0$, and $x$ direction ac driving.
  (a) At $A = 0.2$, the skyrmion moves $10a$ in the $x$ direction
  and $a$ in the y direction during each ac cycle.  
(b) An $R=1/3$ step at $A = 1.0$.
}
\label{fig:24}
\end{figure}

We next consider the case of a fixed dc drive of $F^{D} =1.0$
and changing ac drive amplitude $A$
in a system with $x$ direction ac driving at
$\alpha_{m}/\alpha_{d} = 0.45$.
In Fig.~\ref{fig:23}(a) we plot
$\langle V_{||}\rangle$ and $\langle V_{\perp}\rangle$
versus $A$
and in Fig.~\ref{fig:23}(b) we show the corresponding
$\theta_{sk}$ versus $A$.
When
$A = 0.0$, the skyrmion motion is locked along the $x$ direction, giving
$\theta_{sk} = 0$.
As $A$ increases,
$\langle V_{||}\rangle$ remains fairly constant due to the fixed value of
$F^{D}$,
but small cusps are present which are correlated with
a series of increasing steps in $\langle V_{\perp}\rangle$.
The steps in 
$\langle V_{\perp}\rangle$
produce a series
of steps in
$\theta_{sk} = \arctan(R)$
at $R = 0$, 1/10, 1/6, 1/5, and a small step near $1/4$.
There are extended
steps for $R = 1/3$, $3/7$, and $1/2$.
In general, we find that the magnitude of the
Hall angle
increases with increasing $A$.
In
Fig.~\ref{fig:24}(a) we illustrate
the trajectories for the system in Fig.~\ref{fig:23}
at $A = 0.2$ on the $R = 1/10$ locking step,
where the skyrmion moves $10a$ in the $x$ direction and
$a$ in the $y$ direction during each ac drive cycle.
Figure~\ref{fig:24}(b) shows the same system on the
$R = 1/3$  step at $A = 1.0$,
where the orbit jumps between two different paths
to produce
the $1/3$ ratio.

\begin{figure}
\includegraphics[width=3.5in]{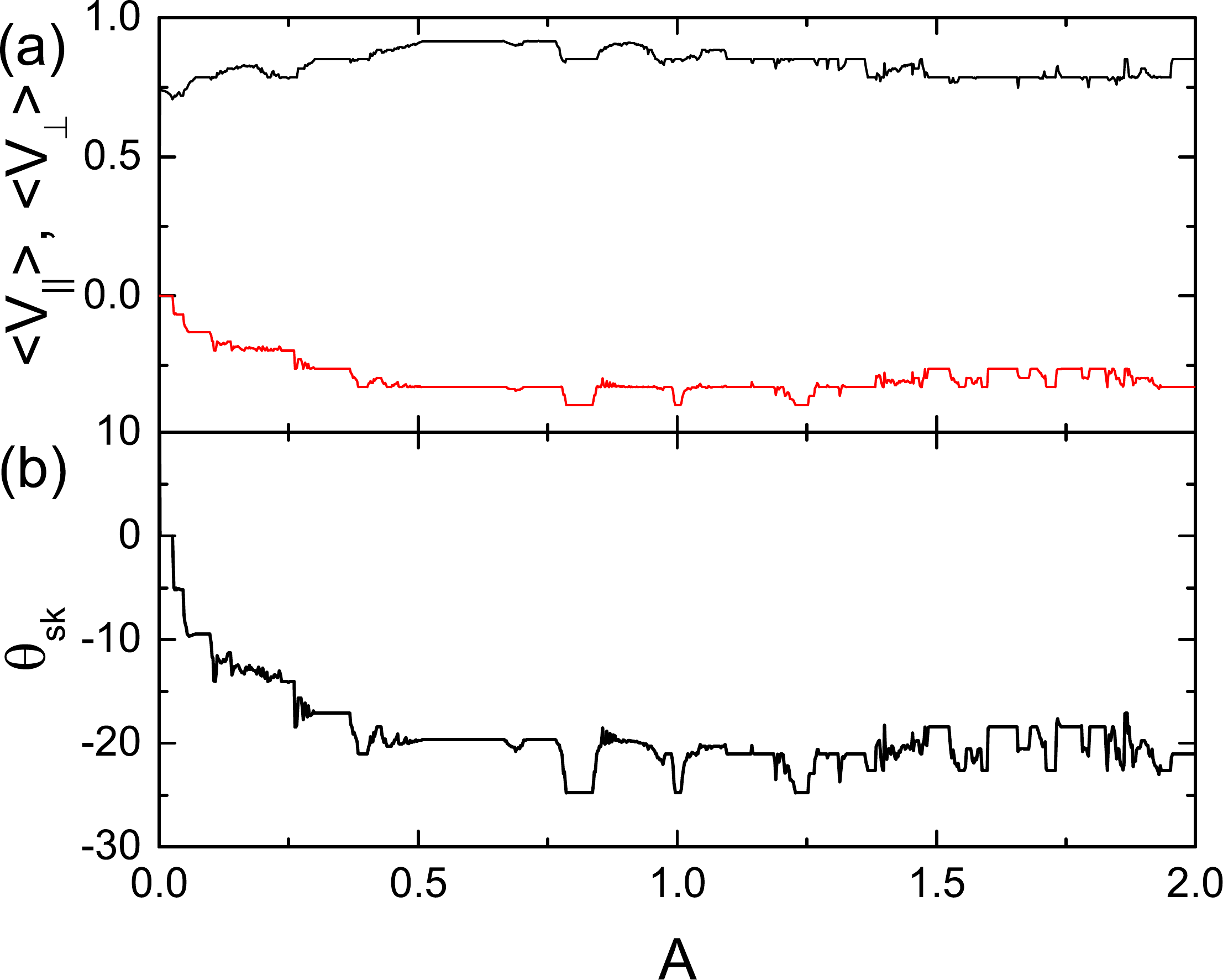}
\caption{
  (a) $\langle V_{||}\rangle$ (black) and $\langle V_{\perp}\rangle$ (red) vs $A$
  for a system with $\alpha_m/\alpha_d=0.45$, $F^D=1.0$,
  and
  $y$ direction ac driving.
  (b) The corresponding $\theta_{sk}$ vs $F^{D}$.
  Here the Hall angle increases in magnitude in a series of steps with increasing
$A$.}
\label{fig:25}
\end{figure}

\begin{figure}
\includegraphics[width=3.5in]{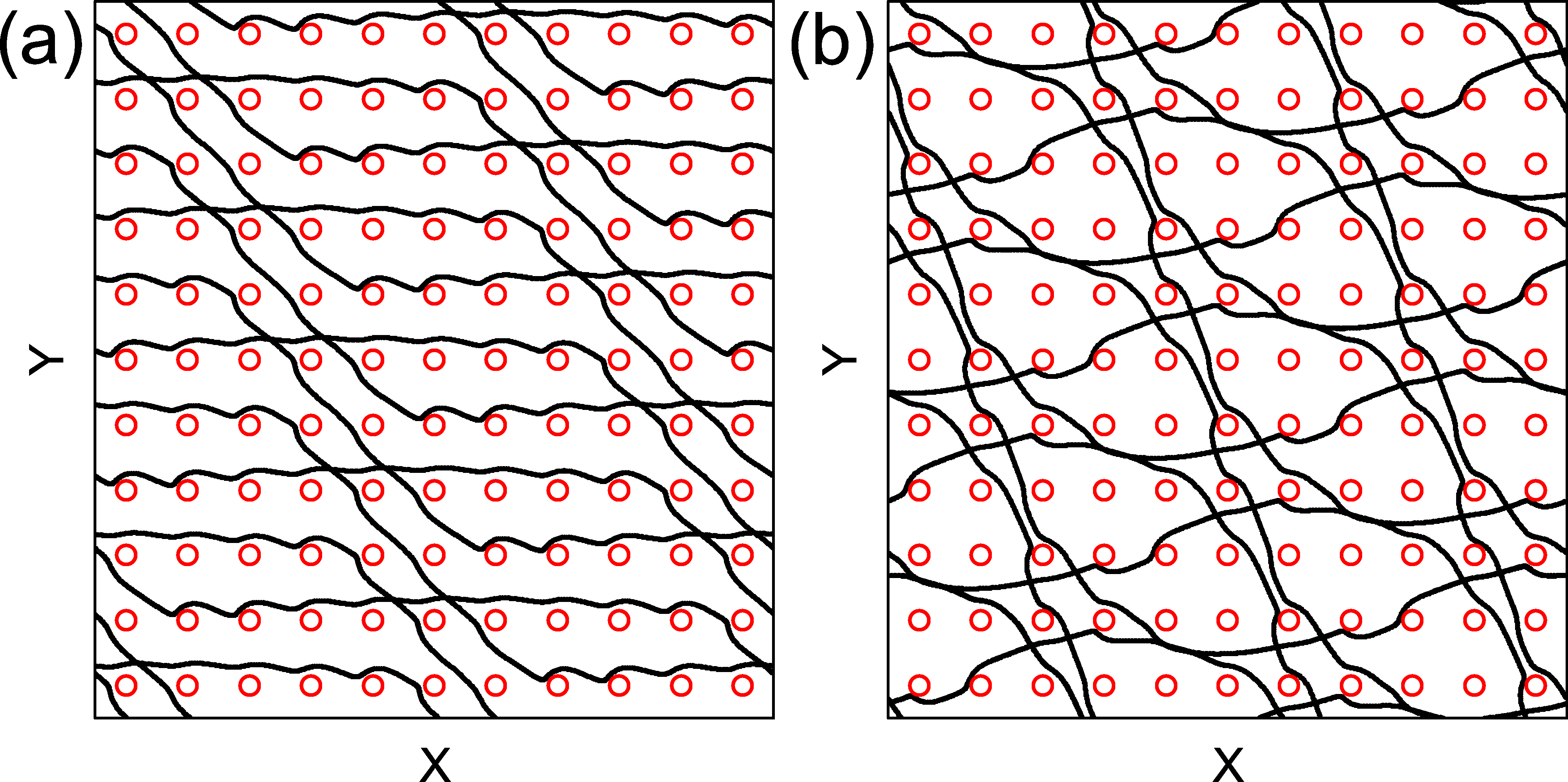}
\caption{ 
  Skyrmion trajectory (black line) and obstacle locations (red circles)
  for the system in Fig.~\ref{fig:25} with $\alpha_m/\alpha_d=0.45$,
  $F^D=1.0$,
  and $y$ direction ac driving. (a) $A = 0.34$.
(b) $A = 1.0$.
}
\label{fig:26}
\end{figure}

In Fig.~\ref{fig:25}(a,b) we plot the velocities and
$\theta_{sk}$ versus $A$ for the same system
as in Fig.~\ref{fig:23} but for ac driving in the $y$-direction.
Here the Hall angle is initially zero since the skyrmion motion is locked
along the $x$ direction.
The velocities and skyrmion
Hall angle increase and decrease in a series of jumps
as $A$ is varied.
In Fig.~\ref{fig:26}(a) we illustrate the trajectories at $A = 0.34$ along a
step on which the skyrmion moves $11a$ in $x$ and $4a$ in $y$ during
every ac cycle.   At $A=1.0$ in Fig.~\ref{fig:24}(b),
there is a more complicated orbit along a
step where the skyrmion moves $11a$ in $x$ and $5a$ in $y$ per ac cycle.
We find similar behavior for higher values of $\alpha_{m}/\alpha_{d}$.
These results indicate that the
Hall angle can be controlled by varying A.

\section{Discussion}
In the locked phases, the skyrmions perform quantized motion along
the $x$ and/or $y$ directions.  This suggests
that ac drives could be used to control skyrmion motion
in different types of devices \cite{Fert13}.
Such controlled motion could be applied
to more complex geometries such as rows of pinning or different tailored
geometries.
We expect that similar results would appear in triangular arrays of obstacles,
where the dominant directional locking
angles are $30^\circ$ and $60^\circ$.
Future areas to address include
the role of temperature, where thermal effects could
strongly affect the transition points or
jumps between different locking phases and
could also produce thermal creep \cite{Reichhardt19}.
At higher temperatures, the phase locking
effects would gradually wash away.
We find that the locking effects are most prominent for systems with
repulsive obstacles,
but if attractive obstacles or pinning sites are used
instead, the locking effects persist
but are smaller for
both the directional locking \cite{Vizarim20} and the Shapiro steps.
We model the skyrmions as point particles;
however, actual skyrmions
often have additional internal modes of motion.
These modes could be excited at much higher frequencies
where
they could induce additional locking frequencies.
Such effects could be explored more fully with continuum based simulations
\cite{Leliaert19}.
In this work we have focused on the motion of a single
isolated skyrmion.
If multiple interacting skyrmions
are present, 
additional locking effects could arise
as a result of
emergent soliton dynamics,
which would be most pronounced just
outside of rational filling fractions of
$1/2$ or $1/1$  \cite{Reichhardt17a,Reichhardt18},
where the filling fraction indicates the ratio of the number of skyrmions to the
number of obstacles or pinning sites.
At commensurate fillings, the skyrmion-skyrmion interactions should cancel
and the dynamics should be similar to the single skyrmion case.

Although our results are focused on skyrmions,
similar effects could arise
for particles in effectively 2D systems
where gyroscopic forces can arise, including
active spinners \cite{vanZuiden16,Han17,Reichhardt19a,Reichhardt19aa}
or charged particles in magnetic fields moving over periodic substrates
\cite{Weiss91,Wiersig01,Power17}. 

\section{Summary}
We have numerically examined a skyrmion moving over a
2D periodic array of obstacles under a dc drive with
an additional ac drive applied either parallel or perpendicular to the
dc driving direction.
We find that the Magnus force induces new
types of dynamical locking effects
that are not observed for overdamped systems
with 2D periodic substrates. 
When the ac and dc drives are parallel, the skyrmion exhibits
both Shapiro steps similar
to those observed in the overdamped case
as well as directional locking in which the skyrmion motion
locks to different symmetry directions of the substrate.
The locking is associated with steps or cusps in the velocities as well as
changes in the skyrmion Hall angle.
Under strictly dc driving, the skyrmion Hall angle changes monotonically with
drive, but when ac driving is added, the skyrmion Hall angle 
can both increase and decrease along the locking steps.
For certain ratios of the Magnus force to the damping term,
we find that even though the skyrmion Hall angle is fixed in a particular
direction of motion,
Shapiro steps still appear
in the velocity force curves
either parallel or perpendicular to the dc drive.
At high drives,
the skyrmion Hall angle gradually approaches the intrinsic value
and shows oscillations 
as a function of increasing drive.
In general, cusps in the velocity force curve are indicative of directional locking,
while steps indicate that Shapiro steps are occurring.
When the ac drive is perpendicular to the dc drive,
we generally find an even larger number of
steps in the velocity force curves and the skyrmion Hall angle.
It is also possible
to observe Hall angle overshoots
in which the skyrmion Hall angle locks
to a value that is much larger than
the intrinsic value.
When the dc drive amplitude is fixed,
steps in the Skyrmion Hall angle can occur as function of changing ac
drive amplitude.
For higher Magnus forces, we generally find
that the steps are reduced and there are increased regions of disordered flow. 

\acknowledgments
This work was supported by the US Department of Energy through
the Los Alamos National Laboratory.  Los Alamos National Laboratory is
operated by Triad National Security, LLC, for the National Nuclear Security
Administration of the U. S. Department of Energy (Contract No.~892333218NCA000001).
N.P.V. acknowledges
funding from
Funda\c{c}\~{a}o de Amparo \`{a} Pesquisa do Estado de S\~{a}o Paulo - FAPESP (Grant 2018/13198-7).

\bibliography{mybib}
\end{document}